\documentclass{article}

\usepackage{arxiv}

\usepackage{cite}
\usepackage{times}
\usepackage{subcaption, graphicx}
\usepackage{amssymb, amsmath, amsthm}          
{
	\theoremstyle{plain}

}
\usepackage{graphicx}
\usepackage{inputenc}
\usepackage{multirow}
\usepackage{makecell}
\usepackage{enumerate}
\usepackage{verbatim}
\usepackage{hyperref}
\usepackage{url}
\usepackage{booktabs}
\usepackage{algorithm}
\usepackage{algorithmic}

\title{Multi-agent Continual Coordination via Progressive Task Contextualization}

\author{%
  Lei Yuan\textsuperscript{\rm 1,2}, Lihe Li\textsuperscript{\rm 1}, Ziqian~Zhang\textsuperscript{\rm 1}, Fuxiang~Zhang\textsuperscript{\rm 1,2}, Cong~Guan\textsuperscript{\rm 1}, Yang Yu\textsuperscript{\rm 1,2,}\thanks{Corresponding Author}\\
  \textsuperscript{\rm 1} National Key Laboratory for Novel Software Technology, Nanjing University, Nanjing, China \\
  \textsuperscript{\rm 2} Polixir.ai\\
  \texttt{\{yuanl, lilh, zhangzq, zhangfx, guanc\}@lamda.nju.edu.cn, yuy@nju.edu.cn}
}

\date{}
\begin{document}

\maketitle

\begin{abstract}
Cooperative Multi-agent Reinforcement Learning (MARL) has attracted significant attention and played the potential for many real-world applications. Previous arts mainly focus on facilitating the coordination ability from different aspects (e.g., non-stationarity, credit assignment) in single-task or multi-task scenarios, ignoring the stream of tasks that appear in a continual manner. This ignorance makes the continual coordination an unexplored territory, neither in problem formulation nor efficient algorithms designed. Towards tackling the mentioned issue, this paper proposes an approach \textbf{M}ulti-\textbf{A}gent \textbf{C}ontinual Coordination via \textbf{Pro}gressive Task Contextualization, dubbed \textbf{MACPro}. The key point lies in obtaining a factorized policy, using shared feature extraction layers but separated independent task heads, each specializing in a specific class of tasks. The task heads can be progressively expanded based on the learned task contextualization. Moreover, to cater to the popular CTDE paradigm in MARL, each agent learns to predict and adopt the most relevant policy head based on local information in a decentralized manner. We show in multiple multi-agent benchmarks that existing continual learning methods fail, while MACPro is able to achieve close-to-optimal performance. More results also disclose the effectiveness of MACPro from multiple aspects like high generalization ability.
\end{abstract}

\section{Introduction}\label{sec:intro}

Cooperative Multi-agent Reinforcement Learning (MARL) has attracted prominent attention in recent years~\cite{oroojlooy2022review}, and achieved great progress in multiple aspects, like path finding~\cite{sartoretti2019primal},  active voltage control \cite{DBLP:conf/nips/WangXGSG21}, and dynamic algorithm configuration~\cite{xue2022multiagent}. Among the multitudinous methods, researchers, on the one hand, focus on facilitating coordination ability via solving specific challenges, including non-stationarity~\cite{papoudakis2019dealing}, credit assignment~\cite{wang2021towards}, and scalability~\cite{christianos2021scaling}. Other works, on the other hand, investigate the cooperative MARL from multiple aspects, like efficient communication~\cite{zhu2022survey}, zero-shot coordination (ZSC)~\cite{DBLP:conf/icml/HuLPF20}, policy robustness~\cite{guo2022towards}, etc. A lot of methods emerge as promising solutions for different scenarios, including policy-based ones~\cite{maddpg,mappo}, value-based series~\cite{vdn,qmix}, and many other variants, showing remarkable coordination ability in a wide range of tasks like SMAC~\cite{gorsane2022towards}. Despite the great success, the mainstream cooperative MARL methods are still restricted to being trained in one single task or multiple tasks simultaneously, assuming that the agents have access to data from all tasks at all times, which is unrealistic for physical agents in the real world that can only attend to one task at a time.

Continual Reinforcement Learning plays a promising role in the mentioned problem~\cite{khetarpal2022towards}, where the agent aims to avoid catastrophic forgetting, as well as enable knowledge transfer to new tasks (a.k.a. stability-plasticity dilemma~\cite{parisi2019continual}), while maintaining scalable to a large number of tasks. Multiple approaches have been proposed to address one or more of these challenges, including regularization-based ones~\cite{kirkpatrick2017overcoming,DBLP:conf/icml/KaplanisSC19,DBLP:conf/aaai/LecarpentierAAJ21}, experience maintaining techniques~\cite{DBLP:conf/nips/Lopez-PazR17,DBLP:conf/icml/CacciaBCP20}, and task structure sharing categories~\cite{DBLP:conf/l4dc/SodhaniMP022,DBLP:conf/aaai/KesslerPBZR22,gaya2022building}, etc. However, the multi-agent setting is much more complex than the single-agent one, as the interaction among agents might cause additional considerations~\cite{zhang2021multi}. Also, coordinating with multiple teammates is proved intrinsically tough~\cite{mirsky2022survey}. Previous works model this problem as multi-task~\cite{DBLP:conf/icml/HuLPF20} or just uni-modal coordination among teammates~\cite{DBLP:conf/icml/NekoeiBCC21}. In light of the significance and ubiquity of cooperative MARL, it is thus imperative to consider the continual coordination in both the problem formulation and the algorithm design to tackle this issue. 

In this work, we develop such a continual coordination framework in cooperative MARL where tasks appear sequentially. Concretely, we first develop a multi-agent task context extraction module, where information of each state in a specific task is extracted and integrated by a product-of-expert (POE) mechanism into a latent space to capture the task dynamic information, and a contrastive regularizer is further applied to optimize the learned representation, with which similar tasks representation are pulled together while dissimilar ones are pushed apart. Afterward, an expandable multi-head policy architecture whose separate independent heads are synchronously expanded with the newly instantiated context, along with a carefully designed shared feature extraction module. Finally, considering the popular CTDE (Centralized Training with Decentralized Execution) paradigm in mainstream cooperative MARL, we leverage the local information of each agent to approximate the policy head selection process via policy distillation in the centralized training process, with which agents can select the most optimal ones to coordinate with other teammates in a decentralized manner.

For the evaluation of the proposed approach, MACPro, we conduct extensive experiments on various cooperative multi-agent benchmarks in the continual setting, including level-based foraging (LBF)~\cite{lbf}, predator-prey (PP)~\cite{maddpg}, and the StarCraft Multi-Agent Challenge benchmark (SMAC)~\cite{pymarl}, and compare MACPro against previous approaches, strong baselines, and ablations. Experimental results show that MACPro considerably improves upon existing methods. More results demonstrate its high generalization ability and its potential to be integrated with different value-based methods to enhance their continual learning ability. Visualization experiments provide additional insight into how MACPro works.

\section{Related Work}
\textbf{Cooperative Multi-agent Reinforcement Learning}
Many real-world problems are made up of multiple interactive agents, which could usually be modeled as a Multi-Agent Reinforcement Learning (MARL) problem~\cite{busoniu2008comprehensive,zhang2021multi}. Further, when the agents hold a shared goal, this problem refers to cooperative MARL~\cite{oroojlooyjadid2019review}, showing great progress in diverse domains like path finding~\cite{sartoretti2019primal},  active voltage control~\cite{DBLP:conf/nips/WangXGSG21}, and dynamic algorithm configuration~\cite{xue2022multiagent}, etc. Many methods are proposed to facilitate coordination among agents, including policy-based ones (e.g., MADDPG~\cite{maddpg}, MAPPO~\cite{mappo}, FD-MARL~\cite{wang2022fully}),  value-based series like VDN~\cite{vdn}, QMIX~\cite{qmix}, Linda~\cite{cao2021linda}, or other techniques like transformer~\cite{wen2022multiagent}, these approaches, have demonstrated remarkable coordination ability in a wide range of tasks (e.g., SMAC~\cite{pymarl}, Hanabi~\cite{mappo}, GRF~\cite{wen2022multiagent}). Besides the mentioned approaches and the corresponding variants, many other methods are also proposed to investigate the cooperative MARL, including efficient communication~\cite{zhu2022survey} to relieve the partial observability caused by decentralized policy execution, policy deployment in an offline manner~\cite{zhang2023discovering}, model learning in MARL~\cite{wang2022model}, policy robustness when some perturbations exist~\cite{guo2022towards}, and training paradigm like CTDE (centralized training with decentralized execution)~\cite{DBLP:conf/atal/LyuXDA21}, ad hoc teamwork~\cite{mirsky2022survey}, etc. 

Despite the mentioned progress, the vast majority of current approaches either focus on training the MARL policy on a single task, or the multi-task setting where all tasks appear simultaneously, lacking the attention to the continual coordination problem. In these methods, MRA~\cite{zhang2021learning} focuses on creating agents that generalize across population-varying Markov games, proposing meta representations for agents that explicitly model the game-common and game-specific strategic knowledge. MATTAR~\cite{qin2022multi} assumes there are some basic tasks, training with which can accelerate the training process in other similar tasks, and develops a multi-agent multi-task training framework. TrajeDi~\cite{DBLP:conf/icml/HuLPF20} and some variants (or improved versions) like MAZE~\cite{xue2022heterogeneous}, concentrate on coordinating with different teammates or even unseen ones like a human, these methods are also under the assumption that we can access all the training tasks all the time. ~\cite{DBLP:conf/icml/NekoeiBCC21} introduces a multi-agent learning testbed that supports both zero-shot and few-shot settings based on Hanabi, but it only considers the uni-modal coordination among tasks, and the experimental results demonstrate methods like VDN~\cite{vdn} trained in the proposed testbed can coordinate well with unseen agents, without any additional assumptions made by previous works.

\textbf{Continual Reinforcement Learning} 
Continual Learning is conceptually related to incremental Learning and Lifelong Learning as they all assume that tasks or samples are presented in a sequential manner~\cite{parisi2019continual,masana2020class,kudithipudi2022biological}. For continual reinforcement learning~\cite{khetarpal2022towards}, EWC~\cite{kirkpatrick2017overcoming} learns new Q-functions by regularizing the $l_2$ distance between the optimal weights of the new task and previous ones. It requires additional supervision information like task changes to update its objective, and then selects a specific Q-function head and a task-specific exploration schedule for different tasks. CLEAR~\cite{DBLP:conf/nips/RolnickASLW19} is a task-agnostic method that does not require task information during the continual learning process, and leverages big experience replay buffers to prevent forgetting. Coreset~\cite{chaudhry2019tiny} prevents catastrophic forgetting by choosing and storing  a significantly smaller subset of the previous task’s data, which is used to rehearse the model during or after finetuning. Some other works like HyperCRL~\cite{DBLP:conf/icra/HuangXBS21}, and \cite{kessler2022surprising} utilize a learned world model to promote continual learning efficiency. Considering the scalability issue along with the task number, CN-DPM~\cite{DBLP:conf/iclr/LeeHZK20} and LLIRL~\cite{DBLP:journals/tnn/WangCD22} decompose the whole task space into several subsets of the data (tasks), and then utilize techniques like Dirichlet Process Mixture or Chinese Restaurant Process to expand the neural network for efficient continual supervised Learning and reinforcement learning tasks, respectively. OWL~\cite{DBLP:conf/aaai/KesslerPBZR22} is a recently proposed approach that learns a multi-head architecture and achieves high learning efficiency, and CSP~\cite{gaya2022building}  incrementally builds a subspace of policies for training a reinforcement learning agent on a sequence of tasks. Other researchers also design  benchmarks like Continual world~\cite{DBLP:conf/nips/WolczykZPKM21}, or baselines~\cite{powers2022cora} to verify the effectiveness of different methods in single-agent reinforcement learning.~\cite{DBLP:conf/icml/NekoeiBCC21} investigate whether agents can coordinate with unseen agents by introducing a multi-agent learning testbed based on Hanabi. Still, it only considers the uni-modal coordination among tasks. Our work takes a further step in this direction for problem formulation and algorithm design.

\section{Problem Formulation} 
This work considers a cooperative multi-agent reinforcement learning problem under partial observation, which can be formalized as a Dec-POMDP~\cite{pomdp}, with tuple $\mathcal{M}=\langle N, \mathcal{S}, \mathcal{A}, \Omega, P, O, R, \gamma \rangle$, where $N = \{1,\cdots,n\}$, $\mathcal{S}$, $\mathcal{A} = \mathcal{A}^1 \times \cdots \times \mathcal{A}^n$, $\Omega$ are the set of agents, states, joint actions, and local observation, respectively. $P: \mathcal{S} \times \mathcal{A} \to \Delta(\mathcal{S})$ stands for the transition probability function, $O: \mathcal{S} \times N \to \Omega$ and $R : \mathcal{S} \times \mathcal{A} \to \mathbb{R}$ are the corresponding  observation function and reward function, and $\gamma \in [0,1)$ is the discounted factor. Multiple interactive agents in a Dec-POMDP coordinate with teammates to complete a task under a share reward $R$, at each time step, agent $i$ receives the local observation $o^i=O(s, i)$ and outputs the action $a^i \in \mathcal{A}^i$. The formal objective of the agents is to maximize the expected cumulative discounted reward $\mathbb{E}[\sum_{t=0}^\infty \gamma^t R(s_t, \pmb{a}_t)]$ by learning an optimal joint policy.

\begin{figure}[]
\centering
\includegraphics[width=0.95\linewidth]{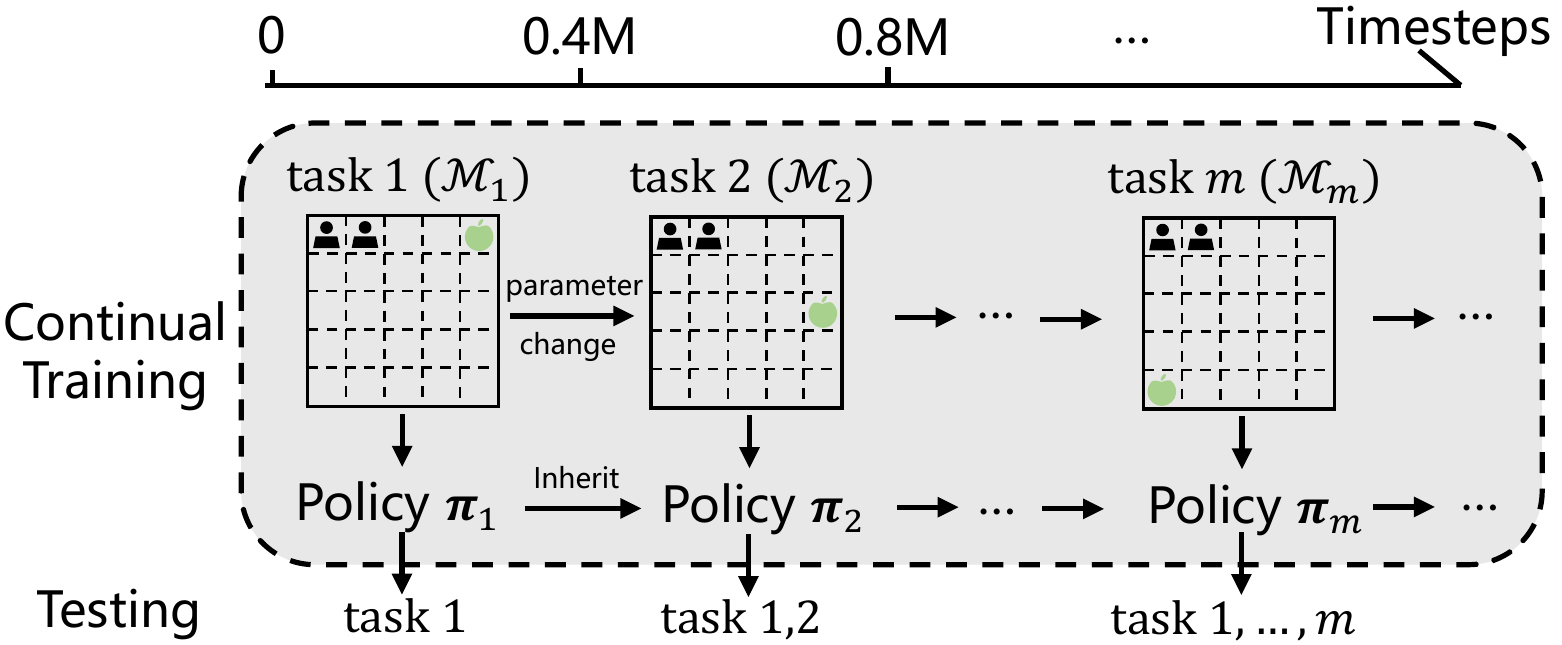}
 \caption{An example of multi-agent continual coordination, where tasks (e.g., the position of food changes in the Level Based Foraging (LBF)~\cite{lbf}) change along with the timeline. We thus need to train a policy $\pmb{\pi}_m$ to solve the concurrent task as well as maintain the knowledge of previous tasks (i.e., avoid catastrophic forgetting) .}
  \label{toy}
\end{figure}
In this work, we focus on a continual coordination problem where agents in a team are exposed to a sequence of (infinite) tasks $\mathcal{Y} = \left(\mathcal{M}_1,\cdots,\mathcal{M}_m, \cdots\right)$. Each task involves a sequential decision making problem and can be formulated as a Dec-POMDP $\mathcal{M}_m=\langle N_m, \mathcal{S}_m, \mathcal{A}_m, \Omega_m, P_m, O_m, R_m, \gamma \rangle$, as shown in Fig.~\ref{toy}. These agents are continually evaluated on all previous tasks (but cannot be trained with these tasks) and the present task. Therefore, the agent's policy needs to transfer to new tasks while maintaining the ability to perform previous tasks. Concretely, agents that have learned $M$ tasks are expected to maximize the MARL objective for each task in $\mathcal{Y}_M = \{\mathcal{M}_1,\cdots,\mathcal{M}_M\}$. We consider the setting where task boundaries are known during the centralized training phase. During the decentralized execution phase, agents cannot access global but only local information to finish the tasks sampled from $\mathcal{Y}_M$.

\begin{figure*}
\setlength{\abovecaptionskip}{0cm}
  \centering
  \includegraphics[scale=0.43]{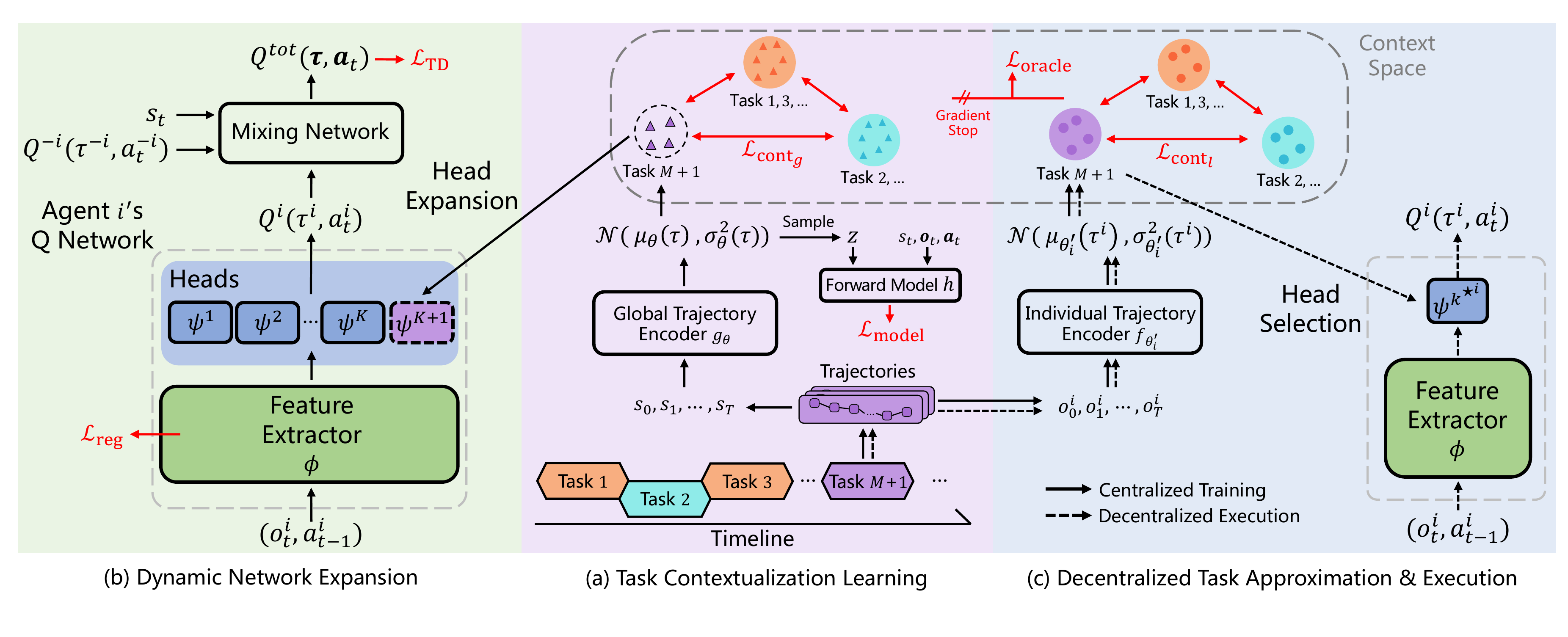}
  \caption{The overall framework of MACPro. (a) We design an efficient multi-agent task contextualization learning module to capture the uniqueness of each emerging task. (b) The training paradigm, including a shared feature extraction part and an adaptive policy heads expansion module based on the learned contexts. (c) Each agent utilizes its local information to approximate the actual task head in a decentralized way.  }
  \label{Structure}
\end{figure*}

\section{Method}
In this section, we will describe the detailed design of our proposed method, MACPro. First, we propose a novel training paradigm, including a shared feature extraction part and an adaptive policy heads expansion module based on the learned contexts (Fig.~\ref{Structure}(a)). Next, we design an efficient multi-agent task contextualization learning module to capture the uniqueness of each emerging task (Fig.~\ref{Structure}(b)). Finally, considering the CTDE property in mainstream cooperative MARL, we train each agent to utilize its local information to approximate the actual task head  (Fig.~\ref{Structure}(c)).


\subsection{Multi-agent Task Contextualization Learning}
In continual reinforcement learning where tasks keep altering sequentially, it is crucial to capture the unique context of each emerging new task. However, the behavioral descriptor of the multi-agent task is much more complex than the single-agent setting due to the interactions among agents~\cite{oroojlooy2022review}. Thus this subsection aims to tackle this issue by developing an efficient multi-agent task contextualization learning module.

Specifically, consider a trajectory $\tau = (s_0, \cdots, s_T)$ with horizon $T$ roll-out by any policies, we utilize a global trajectory encoder $g_\theta$ parameterized by $\theta$ to encode $\tau$ into a latent space. Concretely, the trajectory representation is represented by a multivariate Gaussian
distribution $\mathcal{N}(\mu_\theta(\tau), \sigma^2_\theta(\tau))$ whose parameters are computed by $g_\theta(\tau)$. As the trajectory horizon $T$ may alter for different tasks (e.g., 3m and 5m in SMAC~\cite{pymarl}),  we here apply a transformer~\cite{DBLP:conf/nips/VaswaniSPUJGKP17} architecture (see App.~\ref{Architecture}) to extract feature from each trajectory, thus the latent context of a whole trajectory can be represented as
 $T$ Gaussian distributions $\mathcal{N}(\mu_0, \sigma_0^2), \cdots, \mathcal{N}(\mu_T, \sigma_T^2)$, where $\mathcal{N}(\mu_i, \sigma_i^2)$ stands for the $i^{\text{th}}$ essential parts of the trajectory. Next, considering the importance of different states in a trajectory, we apply the product-of-experts (POE) technique~\cite{hinton2002training} to acquire the joint representation of a trajectory, which is also a Gaussian distribution $\mathcal{N}(\mu_{\theta}(\tau), \sigma^2_{\theta}(\tau))$, where: 
\begin{equation}
    \begin{aligned}
        \mu_\theta(\tau) & = \left(\sum\limits_{t=0}^T \mu_t (\sigma_t^2)^{-1}\right)\left(\sum\limits_{t=0}^T (\sigma_t^2)^{-1}\right)^{-1}, \\
        \sigma^2_\theta(\tau) & = \left(\sum\limits_{t=0}^T (\sigma_t^2)^{-1}\right)^{-1}.
    \end{aligned}
    \label{poe}
\end{equation}
The detailed derivative process between the joint distribution and each single one can be seen in App.~\ref{poeprof}. 

The previous part can obtain representation for each trajectory. Nevertheless, the learned representation lacks any dynamic information about a specific multi-agent task. 
As the difference between any dynamic model lies in transition and reward functions~\cite{luo2022survey}, we here apply a loss function to force the learned trajectory representation to capture the dynamic information of each task. 
Specifically, we learn a context-aware forward model $h$ including three predictors: $h_{s}, h_{o}, h_{r}$ which are responsible to predict the next state, local observations, and reward given the current state, local observations, actions, and task contextualization, respectively:
\begin{equation}
    \begin{aligned}
        \mathcal{L}_{\text{model}} = \mathbb{E}_{\tau\in \mathcal{D}^\prime} \Big[\sum\limits_{t=0}^T & ||h_{s}[s_t, \pmb{o}_t, \pmb{a}_t, z] - s_{t+1}||_2^2 + \\
        & ||h_{o}[s_t, \pmb{o}_t, \pmb{a}_t, z] - \pmb{o}_{t+1}||_2^2 + \\
        & (h_{r}[s_t, \pmb{o}_t, \pmb{a}_t, z] - r_{t})^2\Big],
    \end{aligned}
    \label{model}
\end{equation}
where $z$ is the task contextualization sampled from the joint task distribution, $\mathcal{D}^\prime$ is the replay buffer for task contextualization learning, which stores a small amount of trajectories for each task. 
However, as there are tasks with different correlations, the mentioned optimization object $\mathcal{L}_{\text{model}}$ might be insufficient for differentiable context acquisition. 
Therefore, we apply another auxiliary contrastive loss~\cite{DBLP:conf/cvpr/ChopraHL05} by
pulling together semantically similar data points (positive data pairs) while pushing apart the dissimilar ones (negative data pairs):   
\begin{equation}
\begin{aligned}
    \mathcal{L}_{\text{cont}_g} = & \mathbb{E}_{\tau_j, \tau_k \in \mathcal{D}^\prime} \Big[ \pmb{1}\{y_j = y_k\} D_J(g_\theta(\tau_j) || g_\theta(\tau_k)) + \\
    & \qquad \qquad ~ \pmb{1}\{y_j \not= y_k\} \frac{1}{D_J(g_\theta(\tau_j) || g_\theta(\tau_k)) + \varepsilon}\Big],
\end{aligned}
\label{contrastive loss}
\end{equation}
where $\pmb{1} \{ \cdot \}$ is the indicator function, $y_j$ and $y_k$ are the label(s) of the task(s) from which $\tau_j$ and $\tau_k$ are sampled, respectively, $\varepsilon$ is a small positive constant added to avoid division by zero. $D_J(P||Q) = D_{KL}(P||Q) + D_{KL}(Q||P)$ is the Jeffrey's divergence~\cite{jeffreys1998theory} used to measure the distance between two distributions, and $D_{KL}$ denotes the Kullback-Leibler divergence. 
Thus the overall loss term is: 
\begin{equation}
    \begin{aligned}
        \mathcal{L}_{\text{context}} = \mathcal{L}_{\text{model}} + \alpha_{\text{cont}_g} \mathcal{L}_{\text{cont}_g},
    \end{aligned}
    \label{model loss + contrastive loss}
\end{equation}
where $\alpha_{\text{cont}_g}$ is the  coefficient balancing the loss terms.
\subsection{Adaptive Dynamic Network Expansion}

With the previously learned global trajectory encoder $g_\theta$, we can obtain a unique contextualization for each task. Now, this subsection comes to the design of a context-based continual learning mechanism,  which incrementally clusters a stream of stationary tasks in the dynamic environment into a series of contexts and opts for the optimal policy head from the expandable multi-head neural network.


Formally, for multiple tasks that appear sequentially, we design a policy network consisting of a shared feature extractor $\phi$ with multiple layers of neural network (the index of agent is omitted in this part for simplicity), which can promote knowledge sharing among different tasks. Furthermore, as there may be some multimodal tasks, a single head for all tasks could make the policy overfit to some specific tasks. One way to solve this problem is to learn a customized head for each task like OWL~\cite{DBLP:conf/aaai/KesslerPBZR22}. However, this solution is of poor scalability as the number of heads increases linearly over the number of tasks that could be infinitely many. 
Thus, we develop an adaptive network expansion paradigm based on the similarity between task contextualizations. Specifically, we assume that the agents have already experienced $M$ tasks and have $K$ policy heads $\{\psi^k\}_{k=1}^K$ so far (${\small K \leq M}$). For each head, we store $bs$ trajectories in buffer $\mathcal{D}^\prime$, and we use $g_\theta$ to obtain the corresponding task contextualizations with mean values $\{\{\mu_k^j\}_{j=1}^{bs}\}_{k=1}^{K}$. 


 
When encountering a new task $(M+1)$, we first utilize the feature extractors $\phi$ and all the existing heads $\{\psi^{k}\}_{k=1}^K$ to derive a set of  behavior policies $\{\pmb{\pi}_{k}\}_{k=1}^K$ to collect $bs$ trajectories each on task $(M+1)$, denoted as $\{\{\tau_{k}^j\}_{j=1}^{bs}\}_{k=1}^{K}$. Next, we use $g_\theta$ to derive the mean values $\{\{\mu_k^{\prime j}\}_{j=1}^{bs}\}_{k=1}^{K}$ of their contextualizations 
and calculate the similarities between the existing mean values $\{\{\mu_k^j\}_{j=1}^{bs}\}_{k=1}^{K}$ as follows:
\begin{equation}
    \begin{aligned}
        l & = (l_1, \cdots, l_K),~l^\prime = (l_1^\prime, \cdots, l_K^\prime), \\
        \text{where}~l_k & = \frac{1}{bs} \sum\limits_{j=1}^{bs} ||\mu_k^j - \frac{1}{bs} \sum\limits_{i=1}^{bs} \mu_k^i||_2,\\
        l_k^\prime & = \frac{1}{bs} \sum\limits_{j=1}^{bs} ||\mu^{\prime j}_k - \frac{1}{bs} \sum\limits_{i=1}^{bs} \mu_{k}^i||_2,~k = 1, \cdots, K.
    \end{aligned}
    \label{expansion}
\end{equation}
Here $l$ is the vector describing the dispersion of the $K$ existing contextualizations, and $l^\prime$ is the vector describing the distance between the $K$ new contextualizations and the existing ones. 
Let $k_* = \arg\min_{1 \le k \le K} l^\prime_{k}$, such that the $k_*^{\text{th}}$ pair of existing and new contextualizations are closest among all $K$ pairs. With an adjustable threshold $\lambda_{\text{new}}$, if $l^\prime_{k_*} \le \lambda_{\text{new}} l_{k_*}$, indicating task $(M+1)$ is similar to the task(s) that head $\psi^{k_*}$ takes charge, we thus merge it to this/these learned task(s) and use  the unified head $\psi^{k_*}$ for them.
Otherwise, none of the learned tasks are similar with the new one, a new head $\psi^{K+1}$ is created. This phase processes along with the task sequence, enjoying high scalability and learning efficiency.

The previous part solves the head expansion issue, while a single shared feature extractor may inevitably cause forgetting. We here apply an $l_2$-regularizer to relieve this issue by constraining the parameters of the shared part don't change too drastically when learning task $(M+1)$:
\begin{equation}
    \begin{aligned}
        \mathcal{L}_{\text{reg}} = \sum\limits_{i=1}^n||\phi_i - \phi_i^{M}||_2,
    \end{aligned}
    \label{rec loss}
\end{equation}
where $\phi_i^M$ is the saved snapshot of agent $i$'s feature extractor $\phi_i$ after training on task $M$. As we can apply MACPro to any value-based methods, we thus obtain the temporal difference error $\mathcal{L}_{\text{TD}}$ as 
$\left[r+ \gamma \max_{\boldsymbol{a}^{\prime}}Q^{\rm tot}(s', \boldsymbol{a}^{\prime}; \boldmath \theta^-) - Q^{\rm tot}(s,\boldsymbol{a}; \boldmath \theta)\right]^2 $, where $\boldmath \theta^-$ are parameters of a periodically updated target network. The overall loss term of training agents' policies is defined as follows:
\begin{equation}
    \begin{aligned}
        \mathcal{L}_{\text{RL}} = \mathcal{L}_{\text{TD}} + \alpha_{\text{reg}} \mathcal{L}_{\text{reg}},
    \end{aligned}
    \label{rl loss}
\end{equation}
where $\alpha_{\text{reg}}$ is the coefficient balancing the two loss terms.
\begin{figure*}[!ht]
\setlength{\abovecaptionskip}{0cm}
  \centering
  \includegraphics[scale=0.85]{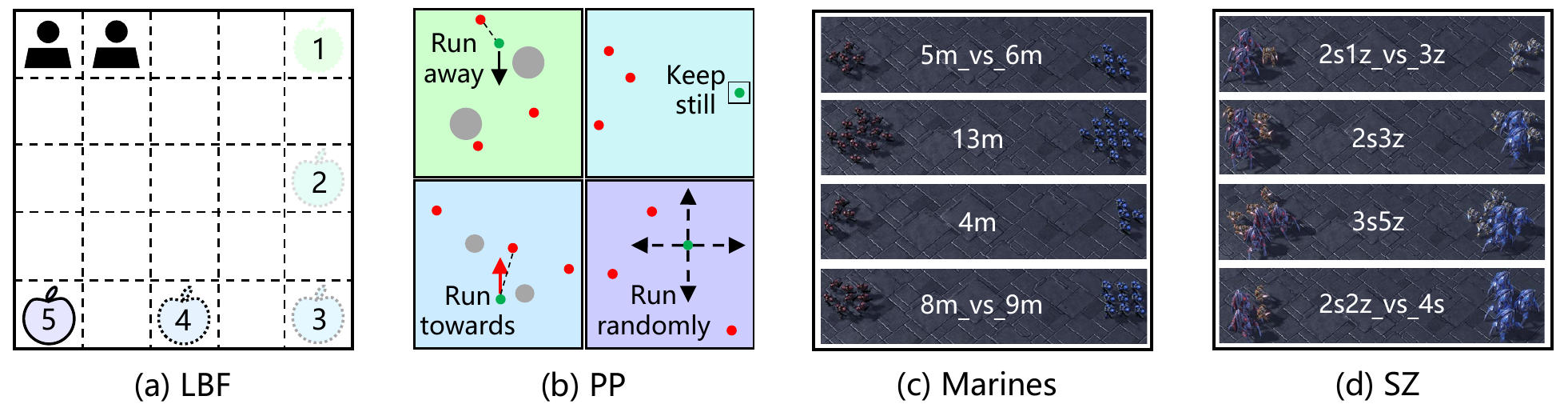}
  \caption{Experimental environments used in this paper. (a) Level-based foraging (LBF), where the position of the food changes in different tasks as indicated by the number on the food. (b) Predator prey (PP), where in different tasks, the position of landmarks, the agents' acceleration, maximum speed, positions, and the fixed heuristic policies the prey uses are different. (c) \& (d) Where Marines and SZ from StarCraft Multi-Agent Challenge (SMAC) involves various numbers and types of battle agents.}
  \label{envs}
\end{figure*}

\subsection{Decentralized Task Approximation}
Although we have obtained an efficient continual learning
approach for any tasks that appear in a sequential way, it is
still far away from the MARL setting, as it requires the trajectory of global states to obtain the task representation, while agents in a MARL system can only acquire its local information.

Towards tackling the mentioned issue, we here develop a distillation solution. Concretely, for agent $i$ with its local trajectory history $\tau^i = (o_0^i, \cdots, o_T^i)$, we design a local trajectory encoder  $f_{\theta_i^\prime}$ that is similar to the global trajectory encoder $g_\theta$. $f_{\theta_i^\prime}$ takes $\tau^i$ as input and outputs $\mathcal{N}(\mu_{\theta_i^\prime}(\tau^i), \sigma^2_{\theta_i^\prime}(\tau^i))$. We thus optimize $f_{\theta_i^\prime}$ by minimizing the Jeffrey's divergence between the distributions: 
\begin{equation}
    \mathcal{L}_{\text{oracle}} = \mathbb{E}_{(\tau, \tau^i) \in \mathcal{D}^\prime}\Big[D_{J}\Big(\overline{g_\theta(\tau)} || f_{\theta_i^\prime}(\tau^i)\Big) \Big],
    \label{approx loss}
\end{equation}
where $\overline{~\cdot~}$ denotes gradient stop, $\tau, \tau^i$ stand for the global and local trajectory of a same task, respectively. To accelerate this learning process and make it consistent with task contextualization learning, we design a local auxiliary contrastive loss:
\begin{equation}
\begin{aligned}
    \mathcal{L}_{\text{cont}_l} = & \mathbb{E}_{\tau_j, \tau_k \in \mathcal{D}^\prime} \Big[ \pmb{1}\{y_j = y_k\} D_J(f_{\theta_i^\prime}(\tau^i_j) || f_{\theta_i^\prime}(\tau^i_k)) + \\
    & \qquad \qquad ~ \pmb{1}\{y_j \not= y_k\} \frac{1}{D_J(f_{\theta_i^\prime}(\tau^i_j) || f_{\theta_i^\prime}(\tau^i_k)) + \epsilon}\Big].
    \label{individual contrastive loss}
\end{aligned}
\end{equation}
The overall loss term of this part is:
\begin{equation}
    \begin{aligned}
        \mathcal{L}_{\text{approx}} = \mathcal{L}_{\text{oracle}} + \alpha_{\text{cont}_l} \mathcal{L}_{\text{cont}_l},
    \end{aligned}
    \label{oracle loss + contrastive loss}
\end{equation}
where $\alpha_{\text{cont}_l}$ is the  coefficient balancing the loss terms.

During the decentralized execution phase, agents firstly roll-out $P$ episodes to probe the environment. 
Concretely, for each probing episode $p (p=1,\cdots, P)$, agents randomly choose one policy head to interact with the evaluating task to collect trajectory $\tau^i_p$, and calculate the mean value $\mu_{\theta_i^\prime}(\tau^i_p)$ of the trajectory representation $f_{\theta_i^\prime}(\tau^i_p)$. Finally, each agent $i$ selects the most optimal task head via comparing the distance with the $K$ existing task contextualization as follows: 
\begin{equation}
    \begin{aligned}
        {k^\star}^i = 
        \underset{1 \le k \le K}{\mathrm{argmin}}\,
        \underset{1 \le p \le P}{\mathrm{min}}\,
        ||\mu_{\theta_i^\prime}(\tau^i_p) - \frac{1}{bs} \sum\limits_{j=1}^{bs} \mu_k^j||_2,
    \end{aligned}
    \label{approximation}
\end{equation}
and use head $\psi^{{k^\star}^i}$ with feature extractor $\phi$ for testing.

\section{Experimental Evaluation}

In this section, we design extensive experiments for the following questions:  1) Can our approach MACPro achieve high continual ability compared to other baselines in different scenarios, and how each component influences its performance?  (Sec.~\ref{results}) ?  2) What task representation is learned by our approach, and how does it influence the continual learning ability (Sec.~\ref{analysis}) ? 3) Can MACPro be integrated into multiple cooperative MARL methods, and how does each hyperparameter influence its performance (Sec.~\ref{generalsensitive}) ?
  \label{main exp}


\subsection{Environments and Baselines} 

For the evaluation benchmarks, we select four multi-agent environments (see Fig.~\ref{envs}). Where Level Based Foraging (LBF)~\cite{lbf} is a cooperative grid world game with agents that are rewarded if they concurrently navigate to the food and collect it. The position of the food changes in different tasks as indicated by the number of the food. Predator Prey (PP)~\cite{maddpg} is another popular benchmark where agents (predators) need to chase the adversary agent (prey) and encounter it to win the game. In different tasks, the position of landmarks, the agents' acceleration, maximum speed, positions, and the fixed heuristic policies the prey uses are different. And Marines and SZ from StarCraft Multi-Agent Challenge (SMAC)~\cite{pymarl}, involving various numbers of the agent. 
\begin{figure*}
\setlength{\abovecaptionskip}{0cm}
  \centering
  \includegraphics[scale=0.550]{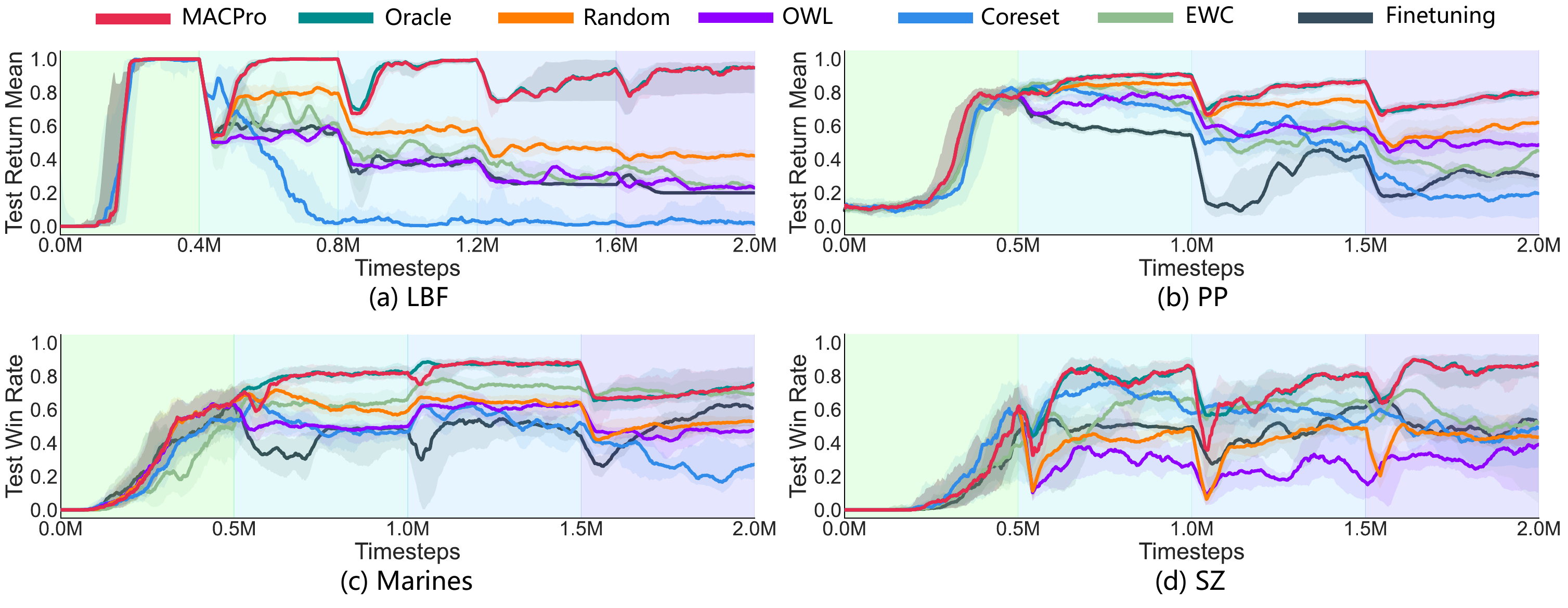}
  \caption{Performance comparison with baselines. Where each task is trained for 400k steps in LBF, 500k steps in other benchmarks, and each plot indicates the average performance across all tasks seen so far.}
  \label{main exp}
\end{figure*}

To evaluate if MACPro can achieve good performance on these benchmarks when different tasks appear continually, we apply it to a popular valued-based method QMIX~\cite{qmix}. Compared baselines include Finetuning, which directly tunes the learned policy on the current task; EWC~\cite{kirkpatrick2017overcoming}, a regularization-based method that constrains the whole agent network from dramatic change; Coreset~\cite{chaudhry2019tiny}, which uses a shared replay buffer over all the tasks so that data on old tasks can rehearse the agents during finetuning on the new task. Also, OWL~\cite{DBLP:conf/aaai/KesslerPBZR22} is included as it is similar to our work but applies the bandit algorithm for head selection. To further study the head selection process, we design Random, with MACPro selecting a head randomly during testing, and Oracle, where MACPro's head selection is based on the ground-truth heads information. More details about benchmarks and baselines can be seen in App.~\ref{baselinebenchmark}.

\subsection{Competitive Results and Ablations} \label{results}
\paragraph{Continual Learning Ability Comparison} 
At first glance, we compare MACPro against the mentioned baselines to investigate the continual learning ability as shown in Fig.~\ref{main exp}. We can find that Finetuning achieves the most inferior performance in different benchmarks, showing that a conventional reinforcement learning training paradigm is improper for continual learning scenarios. Other successful approaches for single agent continual learning, like Coreset, EWC, and OWL, also suffer from performance degradation in the involved benchmarks, demonstrating the necessity of specific consideration for MARL settings. The Oracle baseline, where we give all the ground-truth task identification when testing, can be seen as an upper bound of performance on the related benchmarks, acquiring superiority over all baselines in all benchmarks, demonstrating a multi-head architecture can solve the multi-modal tasks while conventional approaches fail. Our approach MACPro, obtains comparable performance to Oracle, indicating the efficiency of all the designed modules. Random, which selects a head randomly when testing, suffers from terrible performance degradation compared with MACPro and Oracle, showing that the success of MACPro is owing to the appropriate head selection mechanism but not a larger network with multiple heads. 

\begin{figure*}
\setlength{\abovecaptionskip}{0cm}
  \centering
  \includegraphics[scale=0.50]{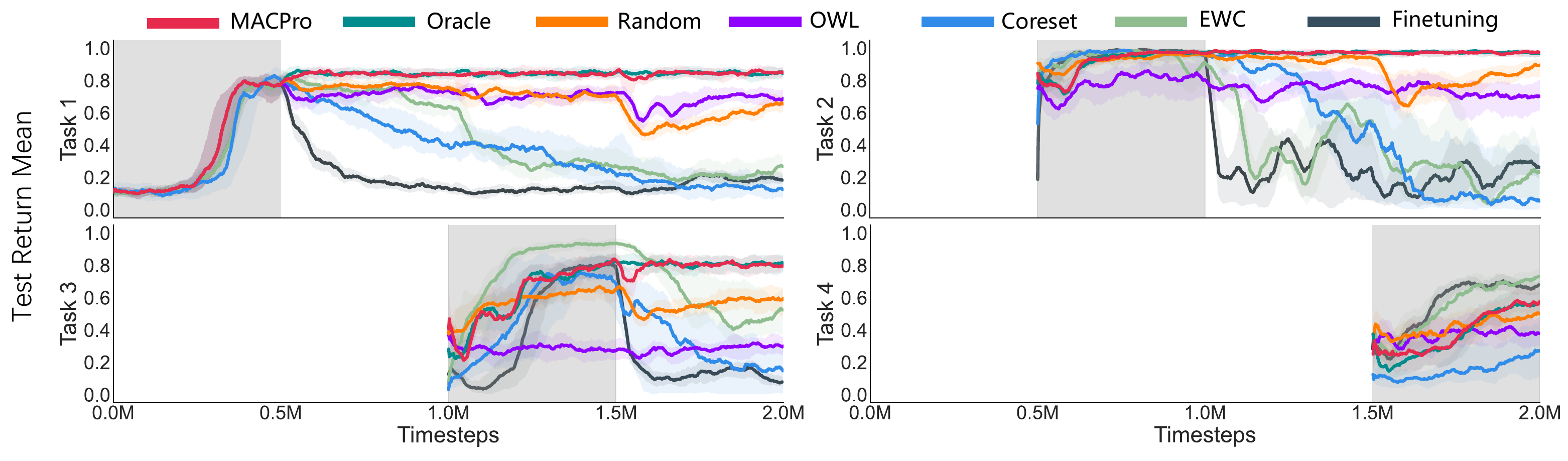}
  \caption{The complete continual learning results on PP. Where five plots for five tasks are displayed. Plot for each task has a grey background, indicating that the agents are training on this task during this period, and there exists blank space in the plots of task $n(n>1)$ because the task has not appeared yet, so we do not test the performance on it until agents have started training on it. For example, task 2 appears at $t=0.4$M and it is trained for $400$k steps, so $t=0.0 \sim 0.4$M is blank and $t=0.4$M $ \sim 0.8$M has grey background.
  }
  \label{exp pp}
\end{figure*}

Furthermore, we display the performance on every single task in PP in Fig~.\ref{exp pp}, We can find that baselines Fientuning, EWC, and Coreset all suffer from performance degradation on one task after training on it, i.e., catastrophic forgetting, demonstrating the necessity of specific consideration for MARL continual learning. other baselines, OWL and Random, fail to choose the appropriate head for testing and does not perform well on all tasks. Learning the new task as quickly as Finetuning without forgetting the old ones, our method MACPro obtains excellent performance. The comparable average performance to Oracle also indicates that MACPro can accurately choose the optimal head for testing. More results can be seen in App.~\ref{moreresult}.
\begin{table*}
\setlength{\abovecaptionskip}{0cm}
\setlength{\tabcolsep}{12pt}
\renewcommand{\arraystretch}{1.5}
    \centering
    \caption{Ablation studies on LBF.}
    \begin{tabular}{l|cccccc}
        \toprule
        Method  & Task 1 & Task 2 & Task 3 & Task 4 & Task 5 & Average \\
        \midrule
        Ours       & \text{$\pmb{1.00 \pm 0.00}$} & \text{$ \pmb{1.00 \pm 0.00}$} & \text{$ 0.80 \pm 0.40$} & \text{$ 0.71 \pm 0.38$} & \text{$\pmb{1.00 \pm 0.00}$} & \text{$ \pmb{0.90 \pm 0.09}$} \\
        W/o model  & \text{$ 0.93 \pm 0.10$} & \text{$ 0.68 \pm 0.46$} & \text{$ 0.67 \pm 0.47$} & \text{$ 0.85 \pm 0.21$} & \text{$ 0.93 \pm 0.10$} & \text{$ 0.81 \pm 0.18$} \\
        W/o cont$_g$  & \text{$\pmb{1.00 \pm 0.00}$} & \text{$ 0.67 \pm 0.47$} & \text{$\pmb{1.00 \pm 0.00}$} & \text{$ \pmb{0.97 \pm 0.03}$} & \text{$ 0.55 \pm 0.41$} & \text{$ 0.84 \pm 0.18$} \\
        W/o POE    & \text{$0.99 \pm 0.01$} & \text{$ 0.86 \pm 0.19$} & \text{$ 0.97 \pm 0.03$} & \text{$ 0.74 \pm 0.17$} & \text{$ 0.69 \pm 0.44$} & \text{$ 0.85 \pm 0.05$} \\
        W/o oracle & \text{$0.21 \pm 0.39$} & \text{$ 0.60 \pm 0.49$} & \text{$ 0.06 \pm 0.12$} & \text{$ 0.24 \pm 0.28$} & \text{$ 0.24 \pm 0.38$} & \text{$ 0.27 \pm 0.07$} \\
        W/o cont$_l$  & \text{$0.98 \pm 0.03$} & \text{$ 0.86 \pm 0.19$} & \text{$0.76 \pm 0.17$} & \text{$ 0.75 \pm 0.20$} & \text{$ 0.97 \pm 0.04$} & \text{$ 0.86 \pm 0.06$} \\
        W/o cont$_{g,l}$  & \text{$ 0.79 \pm 0.40$} & \text{$ 0.99 \pm 0.02$} & \text{$ \pmb{1.00 \pm 0.00}$} & \text{$ 0.34 \pm 0.42$} & \text{$ 0.54 \pm 0.38$} & \text{$ 0.73 \pm 0.12$} \\
        \bottomrule
    \end{tabular}
    \label{ablation}
\end{table*}

\paragraph{Ablation Studies}
As MACPro is composed of multiple components,  we here design ablation studies on benchmark LBF to investigate their impacts. First, for task contextualization learning, we derive \textit{W/o model} by removing the forward model $h$ and its corresponding loss term $\mathcal{L}_{\text{model}}$, and using the contrastive loss only to optimize the global trajectory encoder $g_\theta$. Next, instead of extracting the representation of trajectories with POE, we use the average of the Gaussian distributions generated by the transformer network as the representation, and we call it \textit{W/o POE}.  
Further, we also introduce \textit{W/o oracle}, which has a similar number of parameters as MACPro, to investigate whether the superiority of MACPro over QMIX is due to the increase in the number of parameters.  
Finally, we remove both global contrastive loss $\mathcal{L}_{\text{cont}_g}$ and local contrastive loss $\mathcal{L}_{\text{cont}_l}$ to derive \textit{W/o cont$_{g,l}$}. As shown in Tab.~\ref{ablation}, we can find that when the model loss is removed, \textit{W/o model} suffers from performance degradation in most tasks, indicating the necessity for task representation learning. Furthermore, the POE mechanism also slightly influences the learning performance, demonstrating the special integration of multiple representations of trajectories can facilitate representation learning. Consequently, when removing the oracle loss function, \textit{W/o oracle} sustains great performance degradation, and even fails in task 3, indicating a simply larger network cannot fundamentally improve the performance. We also find  contrastive learning loss has a positive effect on performance.  We further design \textit{W/o cont$_{g}$} and \textit{W/o cont$_{l}$} by setting $\alpha_{\text{cont}_g}=0$ and $\alpha_{\text{cont}_l}=0$ to study the impact of contrastive loss. Both variants suffer from performance degradation, indicating the necessity of contrastive learning.

\begin{figure}[htbp]
\centering
\includegraphics[width=1\linewidth]{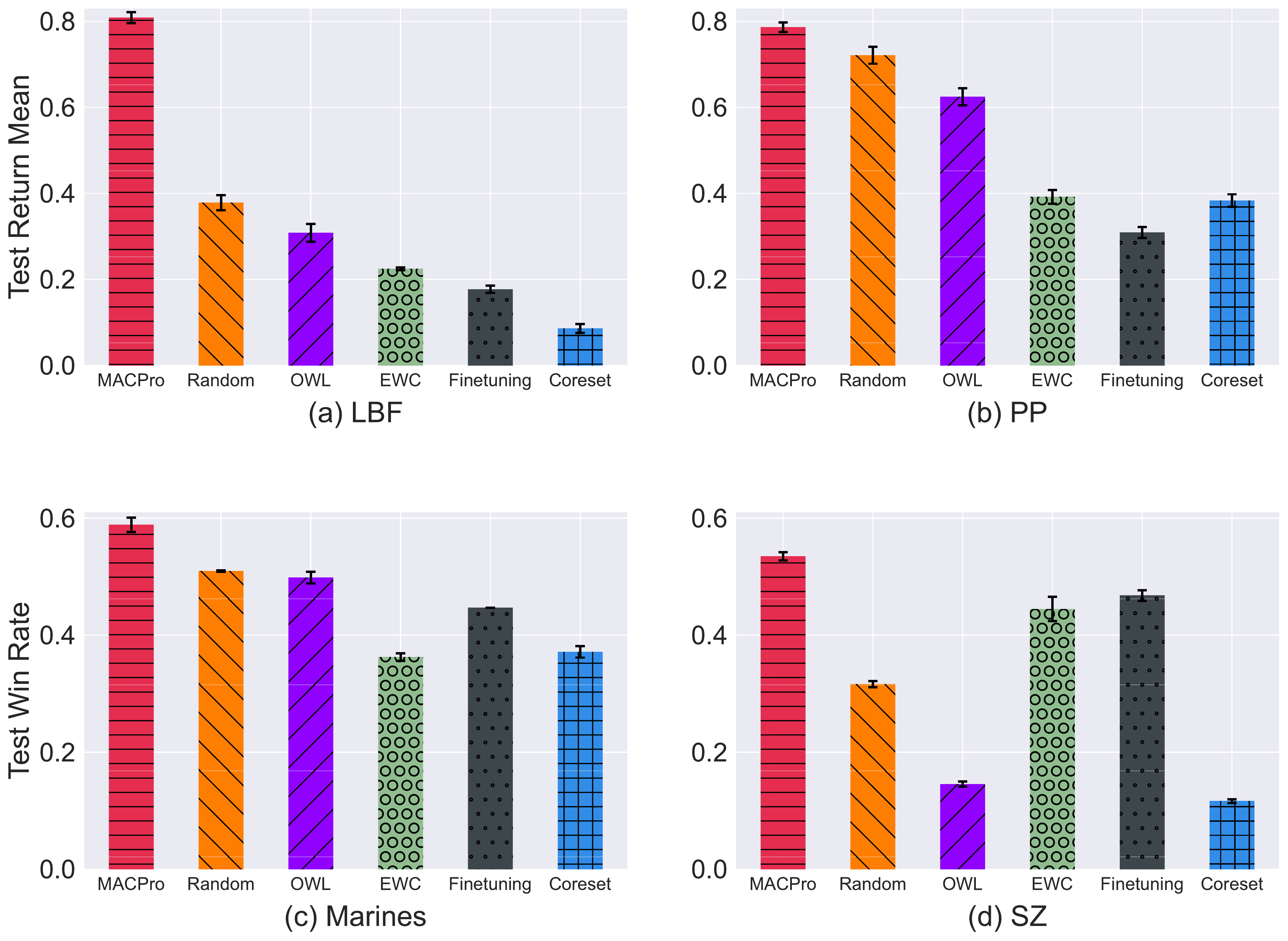}
 \caption{Generalization results.}
  \label{generalization}
\end{figure}

\begin{figure*}[]

\setlength{\abovecaptionskip}{0cm}
  \centering
  \includegraphics[scale=0.48]{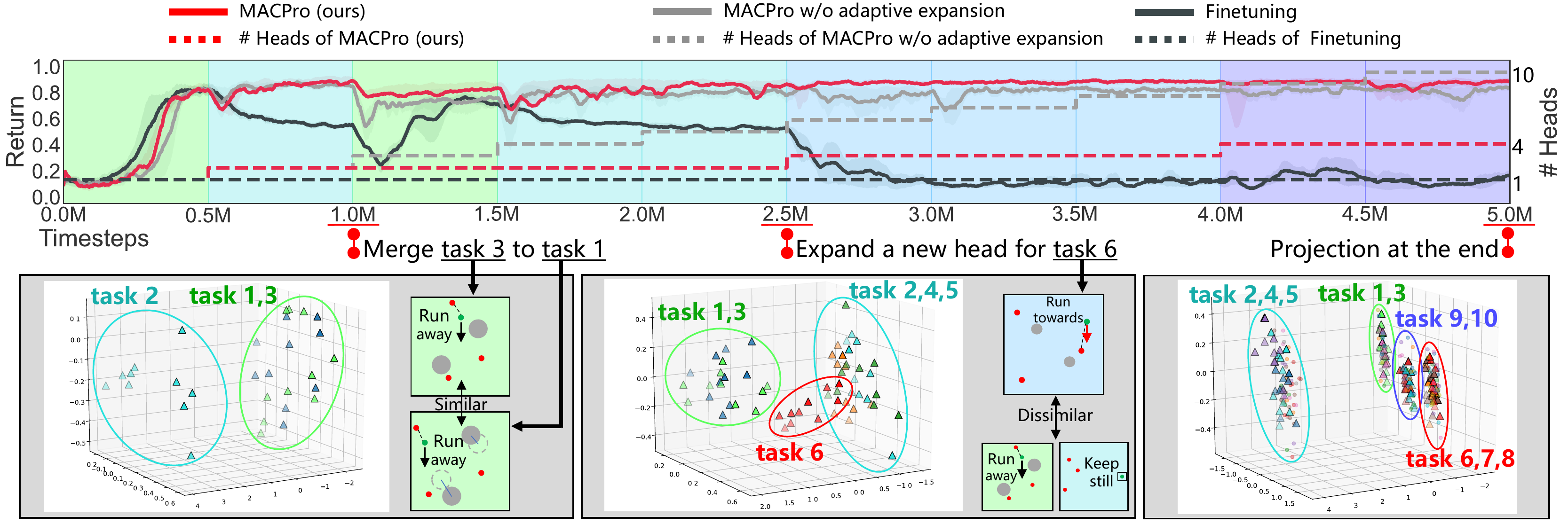}
  \caption{Task contextualization analysis. Where similar tasks have the same background color, e.g., task 1 and task 3 correspond to the green background. When encountering a new task, we sample latent variables generated by  $g_\theta(\tau)$ and apply dimensionality reduction to them by principal component analysis (PCA)~\cite{wold1987principal}, denoted as $\bigtriangleup$.}
  \label{visualization}
\end{figure*}

\paragraph{The Generalization results }

As we focus on training each emerging task sequentially, it produces a significant risk of overfitting. What's more, the ultimate goal of continual learning agents is to not only perform well on seen tasks, but also utilize the learned experiences to complete future unseen tasks. Here, we design experiments to test the generalization ability of MACPro compared with multiple baselines. Concretely, we design 20 additional tasks (details can be seen in App.~\ref{baselinebenchmark}) for each benchmark that agents have not encountered before to conduct zero-shot experiments. 
As shown in Fig.~\ref{generalization}, MACPro demonstrates the most superior performance compared to the multiple compared baselines, indicating that it has strong generalization ability due to the multi-agent task contextualization learning module and the decentralized task approximation procedure. Note that the baseline Oracle is not tested here because there is no ground-truth head selection on unseen tasks.

\subsection{Task Contextualization Analysis} \label{analysis}

Then, we visualize the development of continual learning performance, along with changes in task representation and factored heads, to demonstrate how our method works. Concretely, we build a task sequence with 10 tasks of benchmark PP. As shown in Fig.~\ref{visualization},  when $t=1.0$M, the incoming task 3 is similar to task 1, and their latent variables are distributed in the same area (the green ellipse). Task 3 shares the same head as task 1, leading to an unchanged number of task heads. When a dissimilar task is encountered at $t=2.5$M, none of the learned tasks are similar to the incoming task 6. The latent variables of task 6 are distributed in a new area (the red ellipse), and MACPro does expand a new head accordingly. This process continuously proceeds, until the learning procedure ends at $t=5.0$M, when the latent variables of all ten tasks are distributed in four separate clusters, and MACPro has four heads, respectively. The latent variables of the representations $f_{\theta^\prime_i}(\tau^i)$ encoded by individual trajectory encoders, denoted as $\circ$, are also displayed (we omit them in the first two 3D figures for simplicity). It shows that the representations learned by $f_{\theta^\prime_i}(\tau^i)$ is close to $g_\theta(\tau)$, enabling accurate decentralized task approximation and good performance.

Consequently, we can further find the learning curve in the top row of Fig.~\ref{visualization}, along with the number of separate heads that changes according to the corresponding task representations. We also compare two extreme-case methods, where Finetuning holds a single head for all tasks, enjoying high scalability but strong catastrophic forgetting. On the contrary, MACPro w/o adaptive expansion maintains one head for each task and can achieve high learning efficiency, but the heads' storage cost may impede it when facing a large number of tasks. Our method MACPro, achieves comparable or even better learning ability but consumes fewer heads, showing  high learning efficiency and scalability.

\subsection{Integrative Abilities and  Sensitive Studies} \label{generalsensitive}

MACPro is agnostic to specific value-based cooperative MARL methods. Thus we can use it as a plug-in module and integrate it with existing MARL methods like VDN~\cite{vdn},  QMIX~\cite{qmix}, and QPLEX~\cite{qplex}. As shown in
Tab.~\ref{Integrative}, when integrating with MACPro, the performance of the baselines vastly improves, indicating that MACPro has high generality ability for different methods to facilitate continual learning ability.

\begin{figure}[htbp]
\centering
\includegraphics[width=1\linewidth]{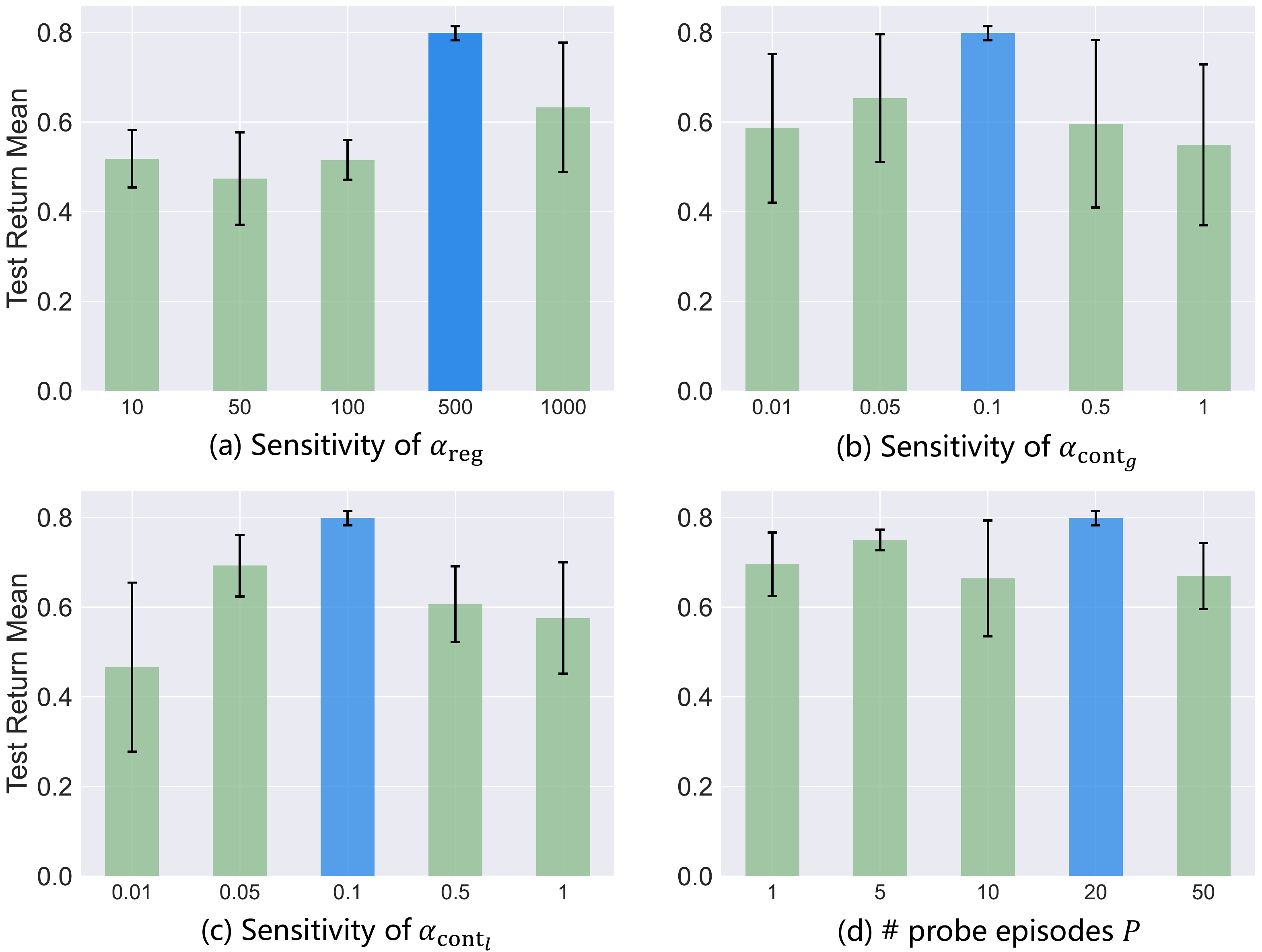}
  \caption{Test results of parameter sensitivity studies.}
  \label{sensitivity main}
\end{figure}

As MACPro includes multiple hyperparameters, here we conduct experiments on benchmark PP to investigate how each one influences the continual learning ability. First, $\alpha_{\text{reg}}$ controls the extent of restriction on changing the parameters of the shared feature extractor $\phi_i$. If it is too small, the dramatic change of $\phi_i$'s parameters may induce severe forgetting. On the other hand, if it is too large, agents remember the old task at the expense of not learning the new task. We thus find each hyperparameter via grid-search.  As shown in Fig.~\ref{sensitivity main} (a), we can find that $\alpha_{\text{reg}}=500$ is the best choice in this benchmark. 
Furthermore, another adjustable hyperparameter $\alpha_{\text{cont}_g}$ influences the training of global trajectory encoder $g_\theta$ in multi-agent task contextualization learning. Fig.~\ref{sensitivity main} (b) shows that $\alpha_{\text{cont}_g}=0.1$ performs the best. 
In decentralized task approximation, $\alpha_{\text{cont}_l}$ balances the learning of local trajectory encoder $f_{\theta_i^\prime}$. We find that in Fig.~\ref{sensitivity main} (c) $\alpha_{\text{cont}_l} = 0.1$ performs the best. During decentralized execution, agents first probe $P$ episodes before evaluation to derive the task contextualization and select the optimal head. The more episodes agents can probe, the more information about the evaluating task agents can gain. However, setting $P$ to a very large value is not practical. We find in Fig.~\ref{sensitivity main} (d) $P=20$ is enough for accurate task approximation.
\begin{table}
\small
\setlength{\tabcolsep}{2.3pt}
\renewcommand{\arraystretch}{1.4}
\centering
    \caption{Integrative Abilities.}
    \begin{tabular}{c|c|ccc} 
    \toprule
    Envs                 & Method     & VDN   & QMIX   & QPLEX   \\ 
    \midrule
    \multirow{2}{*}{LBF} & W/ MACPro  & $ \pmb{ 0.92 \pm 0.02 }$ & $ \pmb{ 0.90 \pm 0.09 }$ & $ \pmb{ 0.97 \pm 0.03 }$  \\
                         & W/o MACPro & $ 0.21 \pm 0.01 $                         & $ 0.20 \pm 0.00 $                         & $ 0.21 \pm 0.01 $                          \\ 
    \midrule
    \multirow{2}{*}{PP}  & W/ MACPro  & $ \pmb{ 0.62 \pm 0.06 }$ & $ \pmb{ 0.80 \pm 0.02 }$ & $ \pmb{ 0.63 \pm 0.05 }$  \\
                         & W/o MACPro & $ 0.29 \pm 0.05 $                         & $ 0.27 \pm 0.04 $                         & $ 0.30 \pm 0.03 $                          \\
    \bottomrule
    \end{tabular}
    \label{Integrative}
\end{table}

\section{Final Remarks}
Observing the great significance and practicability of continual learning, this work takes a further step towards continual coordination in cooperative MARL. We first formulate this problem, where agents are centralized trained with access to global information, then an efficient task contextualization learning module is designed to obtain efficient task representation, and an adaptive dynamic network expansion technique is applied, we finally design a local continual coordination mechanism to approximate the global optimal task head selection. Extensive experiments demonstrate the effectiveness of our approach. To the best of our knowledge, the proposed MACPro is the first multi-agent continual algorithm capable of multi-agent scenarios, which needs a heuristic-designed environment process. Future work on more reasonable and efficient ways, such as environment automatic generation or applying it to real-world applications would be of great value.

\bibliographystyle{IEEEtran}
\bibliography{ref}


{\Large \textbf{Appendix}}

\renewcommand\thesection{\Alph{section}}
\section{Details About Baselines and Benchmarks} 
\label{baselinebenchmark}
This part gives a detailed description about the relevant baselines, benchmarks, and the three related value-based cooperative MARL methods.
\subsection{Baselines}
\textbf{Finetuning} is a simple method based on a single feature extraction model and policy head to learn a sequence of tasks, ignoring the changes in tasks and directly tuning the learned policy on the current task. However, if the current task is different from the previous ones, the parameters of the policy network would change dramatically to acquire good performance on the current task, thus inducing the phenomenon of catastrophic forgetting.

\textbf{EWC}~\cite{kirkpatrick2017overcoming} is one of the regularization-based approaches to address the catastrophic forgetting problem. Concretely, it tries to maintain expertise on old tasks by selectively slowing down learning on the weights that are important for them. Specifically, the loss function for learning the current task $M$ is
\begin{equation}
    \mathcal{L}(\theta) = \mathcal{L}_M(\theta) + \frac{\lambda}{2} \sum\limits_j  F_j (\theta_j - \theta_{M-1, j})^2,
\end{equation}
where $\mathcal{L}_M(\theta)$ is the loss for task $M$ only, and $F_i$ is the $i^{\text{th}}$ diagonal element of the Fisher information matrix $F$. $\theta_{M-1}$ is the saved snapshot of $\theta$ after training task $M-1$, and $j$ labels each parameter. $\lambda$ is a adjustable coefficient to control the trade-off between the current task and previous ones. In this paper, we set the parameters of the agent's $Q$ network as $\theta$, and calculate the Fisher information matrix $F$ with temporal difference error.

Unlike EWC that constrains the change of network parameters when learning a new task, \textbf{Coreset}~\cite{chaudhry2019tiny}, one of the replay-based methods, prevents catastrophic forgetting by choosing and storing a significantly smaller subset of data of the previous tasks. When learning the current task, the stored data is also utilized for training the policy, which is expected to remember the previous tasks. In this paper, we set the replay buffer to uniformly store trajectories of all seen tasks, including the current one. A small batch of trajectories of one randomly chosen task is sampled from the buffer to train the agents' network.

\textbf{OWL}~\cite{DBLP:conf/aaai/KesslerPBZR22} is a recent approach that learns a multi-head architecture and achieves high learning efficiency when the tasks in a sequence have conflicting goals. Specifically, it learns a factorized policy with a shared feature extractor but separate heads, each specializing in only one task. With a similar architecture to our method MACPro, we can apply it to learn task sequences in a continual manner. During testing, OWL uses bandit algorithms to find the policy that achieves the highest test task reward. However, this strategy could bring performance degradation since agents choose action uniformly at the beginning of the episodes.

\begin{figure}
  \centering
  \includegraphics[scale=0.62]{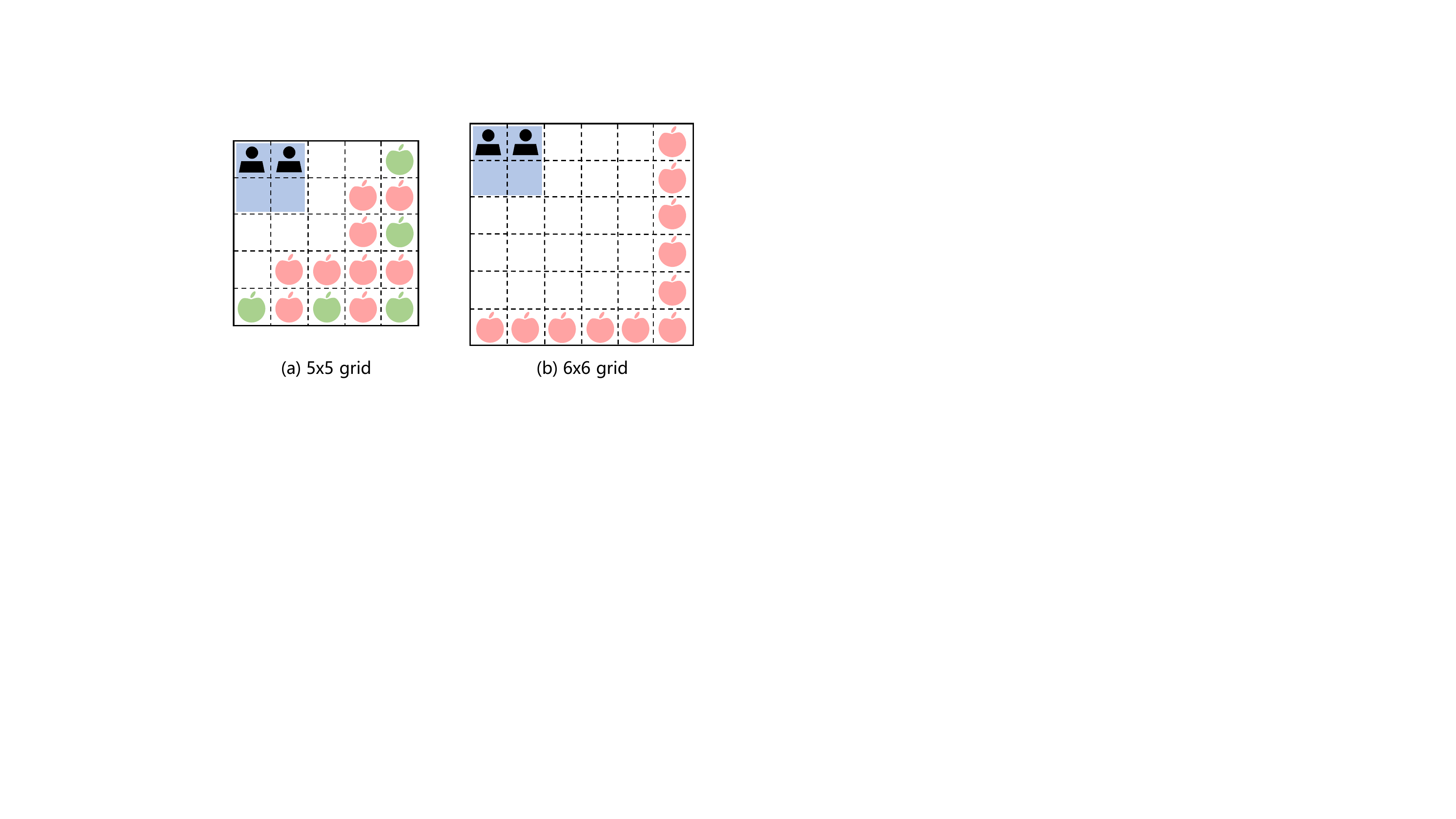}
  \caption{Benchmark LBF used in this paper.}
  \label{env lbf}
\end{figure}

\subsection{Benchmarks}
\begin{figure}
  \centering
  \includegraphics[scale=0.32]{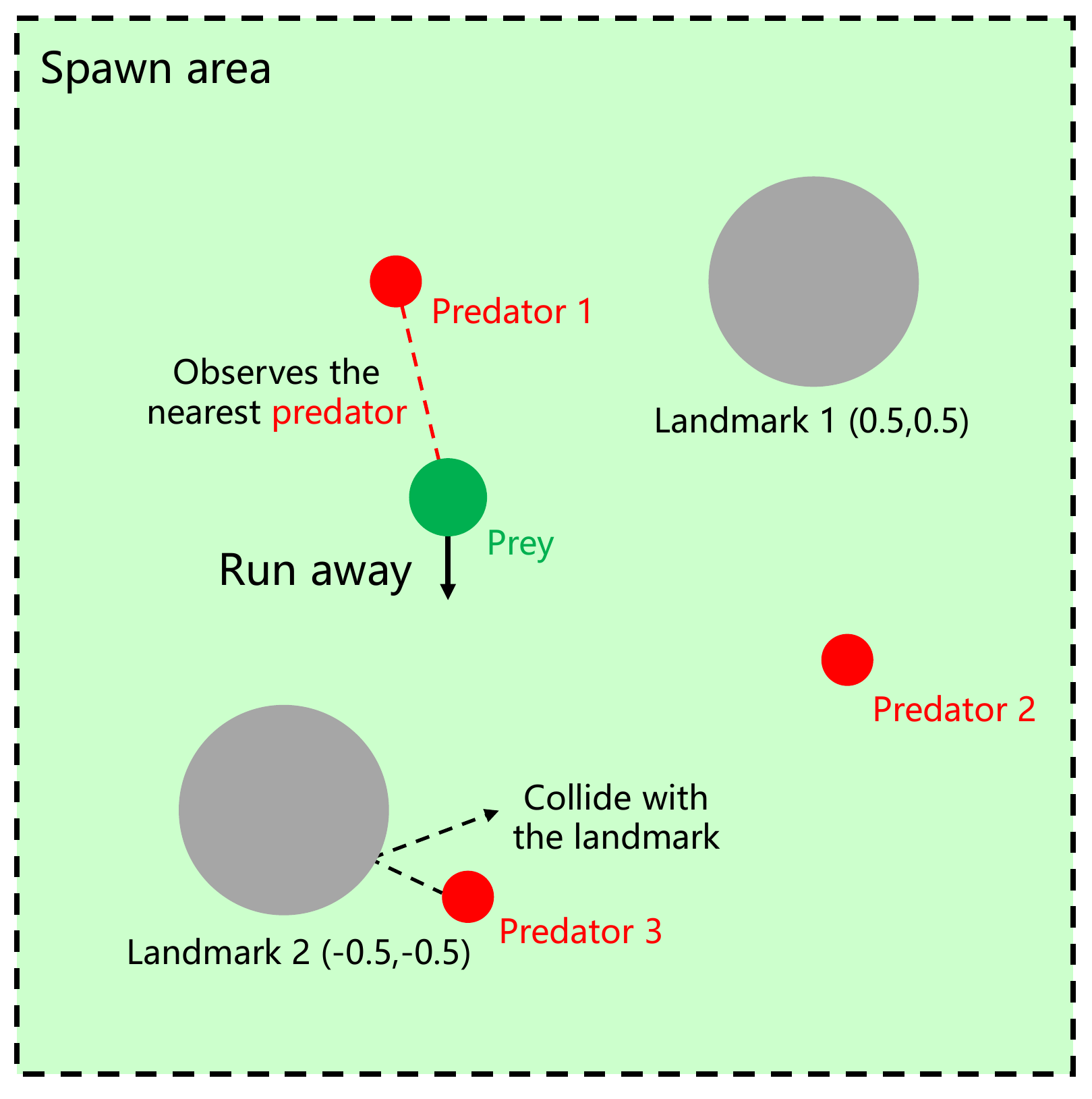}
  \caption{Benchmark PP used in this paper, where the prey's policy is to run in the opposite direction of the nearest predator.}
  \label{env pp}
\end{figure}

We select four multi-agent environments for the evaluation benchmarks. 
\textbf{Level Based Foraging (LBF)}~\cite{lbf} is a cooperative grid world game (see Fig.~\ref{env lbf}). Where the positions of two agents and one food are represented by discrete states, and agents are randomly spawned at cells $(0,0), (0,1), (1,0), (1,1)$. Each agent observes the relative position of other agents and the food, moves a single cell in one of the four directions (up, left, down, right), and gains reward $1$ if and only if both agents navigate to the food and be at a distance of one cell from the food. In continual learning ability comparison, we design 5 tasks in a 5x5 grid, with the food at cell $(0,4), (2,4), (4,4), (4,2), (4,0)$ (green food in Fig.~\ref{env lbf} (a)), respectively. To test the generalization ability of different methods, we further design 20 tasks in both 5x5 and 6x6 grid, and the food position is changed as well  (red food in Fig.~\ref{env lbf} (a)(b)).

\textbf{Predator Prey (PP)}~\cite{maddpg} is another popular benchmark where three agents (predators) need to chase the adversary agent (prey) and collide it to win the game (see Fig.~\ref{env pp}). Here agents and landmarks are represented by circles with different sizes and colliding means circles' intersection. Positions of the two fixed landmarks, positions and speed of the predators and the prey are encoded into continuous states. The predators and the prey can accelerate in one of the four directions (up, left, down, right). In different tasks, the position of landmarks, the predators and the prey's acceleration, maximum speed, and spawn areas, and the fixed heuristic policies the prey uses are different. Specifically, the prey (1) runs in the opposite direction of the nearest predator, (2) stays still at a position far away from the predators, (3) runs towards the nearest predator, and (4) runs in a random direction with great speed. Predators gains reward $1$ if $n$ of them collide with the prey at the same time ($n=1$ in case (1)(2)(4) and $n=2$ in case (3)). In the generalization test, for one original task, we create one corresponding additional task by adding one constant $\xi$ to the original value of different task parameters, including landmark's size, x-coordinate, y-coordinate, predator's and prey's size, acceleration, and maximum speed. We set $ \xi = \pm 0.01, \pm 0.02, 0.03$ on the four original tasks to derive 20 addition tasks to test the generalization ability.
\begin{figure}
  \centering
  \includegraphics[scale=0.25]{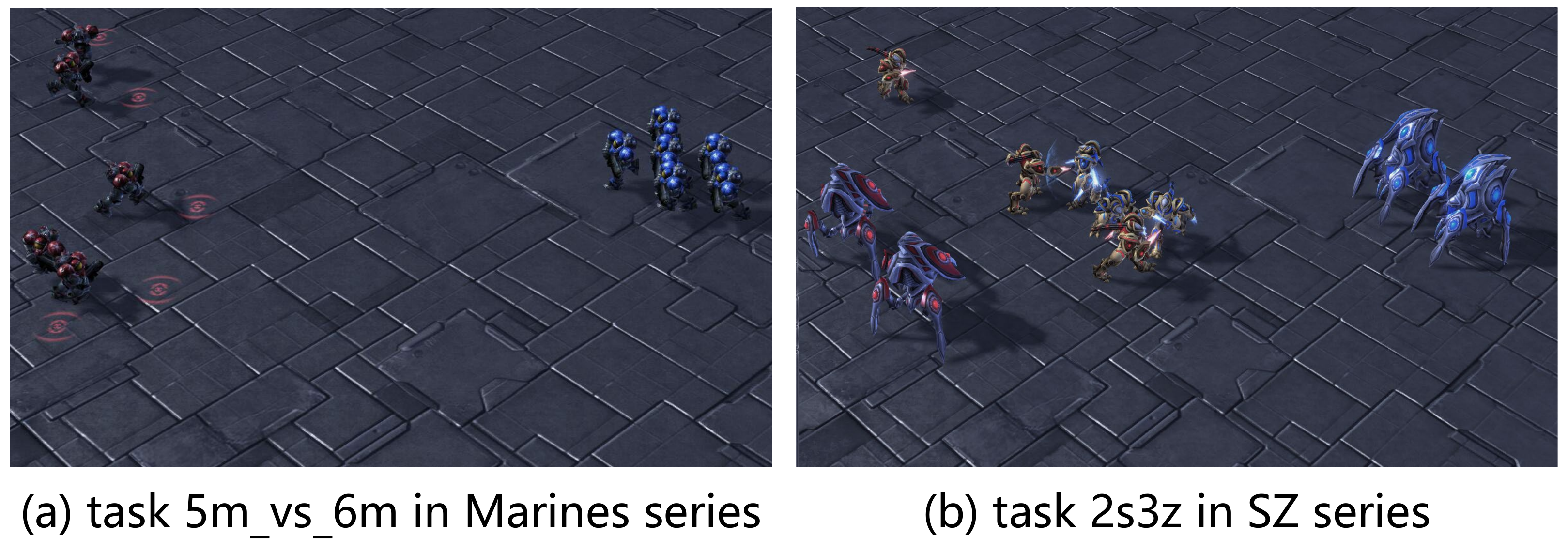}
  \caption{Benchmark Marines and SZ used in this paper.}
  \label{env sc2}
\end{figure}

The other benchmarks are two task series, named \textbf{Marines} and \textbf{SZ} (Fig.~\ref{env sc2}), from StarCraft Multi-Agent Challenge (SMAC)~\cite{pymarl}, involving various numbers of Marines, Stalkers/Zealots in two camps, respectively. The goal of the multi-agent algorithm is to control one of the camps to defeat the other. Agents receive a positive reward signal by causing damage to enemies, killing enemies, and winning the battle. On the contrary, agents receive a negative reward signal when they receive damage from enemies, get killed, and lose the battle. Each agent observes information about the map within a circular area around it, and takes actions, including moving and firing when it is alive. In continual learning ability comparison, the Marines series consists of (1) 5m\_vs\_6m, (2) 13m, (3) 4m, and (4) 8m\_vs\_9m, and the SZ series consists of (1) 2s1z\_vs\_3z, (2) 2s3z, (3) 3s5z, and (4) 2s2z\_vs\_4s, where \textit{m} stands for \textit{marine}, which can attack an enemy unit from a long distance at a time, \textit{s} stands for the \textit{stalker}, which attacks like a marine and has a self-regenerate shield, and \textit{z} stands for \textit{zealot}, which also has a self-regenerate shield but can only attack an enemy unit from a short distance. For the generalization test, we first decrease the default sight range and shoot range by 1 to create four additional tasks for both Marines and SZ. Then, we design scenarios \{3m,
                5m,
                6m,
                7m,
                8m,
                9m,
                10m,
                11m,
                12m,
                4m\_vs\_5m,
                6m\_vs\_7m,
                7m\_vs\_8m,
                9m\_vs\_10m,
                10m\_vs\_11m,
                11m\_vs\_12m,
                12m\_vs\_13m\} for Marines, and scenarios \{1s1z,
              1s2z,
              1s3z,
              2s1z,
              2s2z,
              2s4z,
              3s1z,
              3s2z,
              3s3z,
              3s4z,
              4s2z,
              4s3z,
              4s4z,
              2s2z\_vs\_4z,
              3s3z\_vs\_6s,
              4s4z\_vs\_8z\} for SZ.
If \textit{vs} in the task name, it indicates that the two camps are asymmetric, e.g., in 5m\_vs\_6m, there are 5 marines in our camp and 6 enemy marines. Otherwise, it indicates that the two camps are symmetric, e.g., in 2s3z, there are 2 stalkers and 3 zealots in both camps.

\subsection{Value Function factorization MARL methods}
As we investigate the integrative abilities of MACPro in the manuscript, here we introduce the value-based methods used in this paper, including VDN~\cite{vdn}, QMIX~\cite{qmix}, and QPLEX~\cite{qplex}. The difference among the three methods lies in the mixing networks, with increasing representational complexity. Our proposed framework MACPro follows the \textit{Centralized Training with Decentralized Execution} (CTDE) paradigm used in value-based MARL methods. 

These three methods all follow the Individual-Global-Max (IGM)~\cite{QTRAN} principle, which asserts the consistency between joint and local greedy action selections by the joint value function $Q_{\rm tot}(\boldsymbol{\tau}, \boldsymbol{a})$ and individual value functions $\left[Q_i(\tau^i, a^i)\right]_{i=1}^n$:
\begin{equation}
\begin{aligned}
  & \forall \boldsymbol{\tau} \in \boldsymbol{\mathcal{T}}, \underset{\boldsymbol{a} \in \boldsymbol{\mathcal{A}}}{\arg \max } Q_{\rm tot}(\boldsymbol{\tau}, \boldsymbol{a})= \\
     &\left(\underset{a^{1} \in \mathcal{A}}{\arg \max } Q_{1}\left(\tau^{1}, a^{1}\right), \ldots, \underset{a^{n} \in \mathcal{A}}{\arg \max } Q_{n}\left(\tau^{n}, a^{n}\right)\right).
     \end{aligned}
\end{equation}

\textbf{VDN} \cite{vdn} factorizes the global value function $Q_{\rm tot}^{\text{VDN}}(\boldsymbol{\tau}, \boldsymbol{a})$ as the sum of all the agents' local value functions $\left[Q_i(\tau^i, a^i)\right]_{i=1}^n$:
\begin{equation}
     Q_{\rm tot}^{\mathrm{VDN}}(\boldsymbol{\tau}, \boldsymbol{a})=\sum_{i=1}^{n} Q_{i}\left(\tau^{i}, a^{i}\right).
\end{equation}
\begin{figure}
\centering
\includegraphics[width=0.45\textwidth]{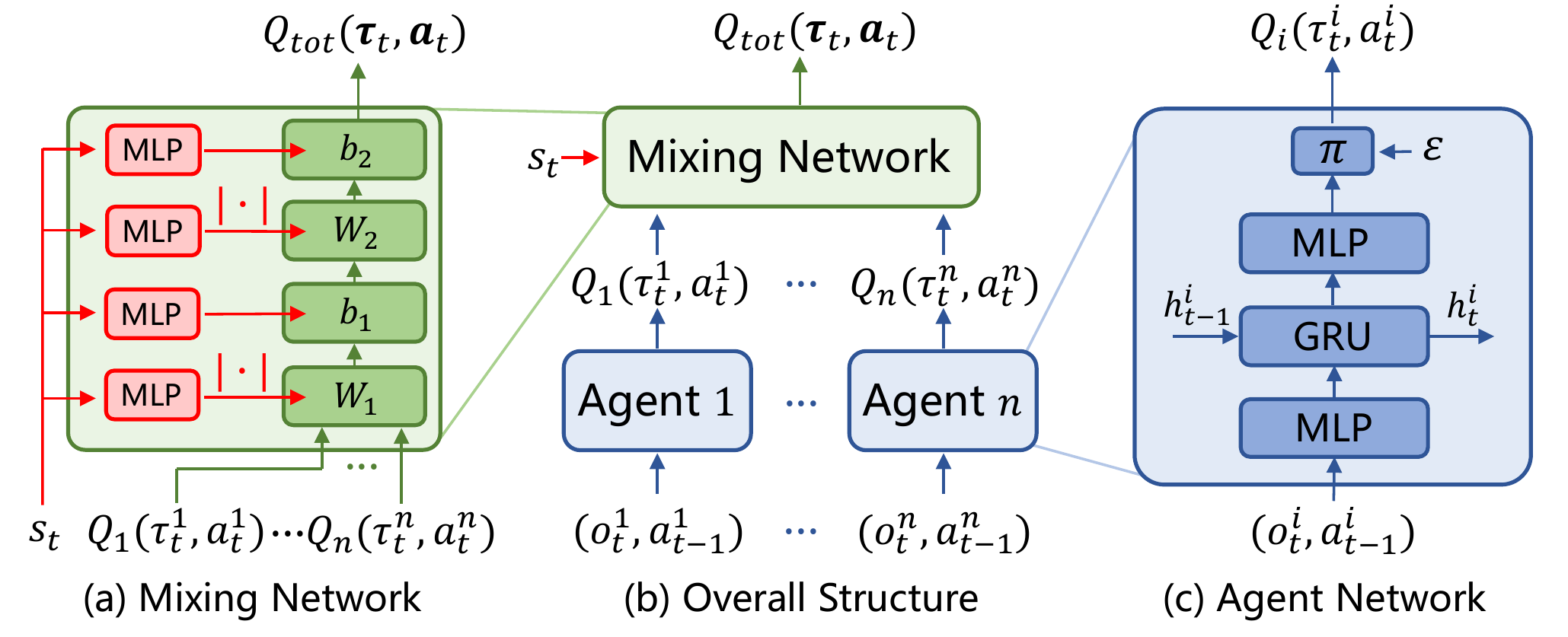}
\caption{The overall structure of QMIX. (a) The detailed structure of the mixing network, whose weights and biases are generated from a hyper-net (red) which takes the global state as the input. (b) QMIX is composed of a mixing network and several agent networks. (c) The detailed structure of the individual agent network. }
\label{fig:QMIX}
\end{figure}
\textbf{QMIX} \cite{qmix} extends VDN by factorizing the global value function $Q_{\rm tot}^{\mathrm{QMIX}}(\boldsymbol{\tau}, \boldsymbol{a})$ as a monotonic combination of the agents' local value functions $\left[Q_i(\tau^i, a^i)\right]_{i=1}^n$:
\begin{equation}
     \forall i \in \mathcal{N}, \frac{\partial Q_{\rm tot}^{\mathrm{QMIX}}(\boldsymbol{\tau}, \boldsymbol{a})}{\partial Q_{i}\left(\tau^{i}, a^{i}\right)}>0.
\end{equation}
\begin{figure*}[!ht]
\setlength{\abovecaptionskip}{0cm}
  \centering
  \includegraphics[scale=0.50]{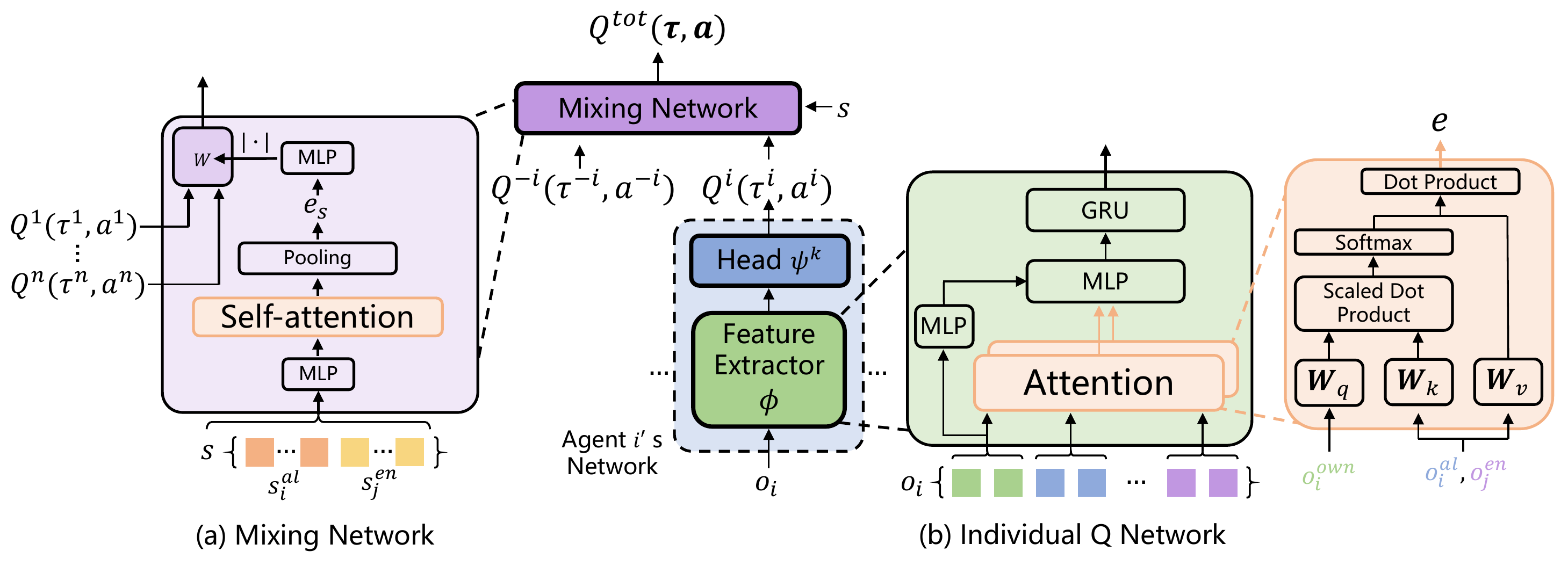}
  \caption{Network architecture used in Marines and SZ.}
  \label{exp pin}
\end{figure*}
We mainly implement MACPro on QMIX for its proven performance in various papers and its overall structure is shown in Fig.~\ref{fig:QMIX}. QMIX uses a hyper-net conditioned on the global state to generate the weights and biases of the local Q-values and uses the absolute value operation to keep the weights positive to guarantee monotonicity.

 The above two structures propose two sufficient conditions of the IGM principle to factorize the global value function but these conditions they propose are not necessary. To achieve a complete IGM function class, \textbf{QPLEX}~\cite{qplex} uses a duplex dueling network architecture by decomposing the global value function as:
\begin{equation}
\begin{aligned}
     &Q_{\rm tot}^{\mathrm{QPLEX}}(\boldsymbol{\tau}, \boldsymbol{a})=V_{\rm tot}(\boldsymbol{\tau})+A_{\rm tot}(\boldsymbol{\tau}, \boldsymbol{a})=\\
     &\sum_{i=1}^{n} Q_{i}\left(\boldsymbol{\tau}, a^{i}\right)+\sum_{i=1}^{n}\left(\lambda^{i}(\boldsymbol{\tau}, \boldsymbol{a})-1\right) A_{i}\left(\boldsymbol{\tau}, a^{i}\right),
     \end{aligned}
\end{equation}
where $\lambda^{i}(\boldsymbol{\tau}, \boldsymbol{a})$ is the weight depending on the joint history and action, $A_{i}\left(\boldsymbol{\tau}, a^{i}\right)$ is the advantage function conditioning on the history information of each agent. QPLEX aims to find the monotonic property between individual Q function and individual advantage function.

\section{The Architecture, Infrastructure, and Hyperparameters Choices of MACPro}
\label{Architecture}
We give detailed description of the network architecture, the overall flow, and the parameters of MACPro here.
\subsection{Network Architecture}
We here give details about multiple neural networks in (1) agent networks, (2) task contextualization learning, and (3) decentralized task approximation.

In benchmark LBF and PP, the number of agents, the dimension of state, observation, and action remains unchanged in different tasks. Specifically, for (1) agent networks, we apply the technique of parameter sharing and design the feature extractor $\phi$ as a 5-layer MLP and a GRU~\cite{cho2014learning}. The hidden dimension is 128 for the MLP and 64 for the GRU. Then, each separated head is a linear layer which takes the output of the feature extractor as input and outputs the Q-value of all actions. For (2) task contextualization learning, we design a global trajectory encoder $g_\theta$ and a context-aware forward model $h$. $g_\theta$ consists of a transformer encoder, a MLP, and a POE module. The 6-layer transformer encoder takes trajectory $\tau = (s_1, \cdots, s_T)$ as input and outputs $T$ 32-dimensional embeddings. Then, the 3-layer MLP transforms these embeddings into means and standard deviations of $T$ Gaussians. Finally, the POE module acquires the joint representation of the trajectory, which is also a Gaussian distribution $\mathcal{N}(\mu_\theta(\tau), \sigma_\theta^2(\tau))$. The context-aware forward model $h$ is a 3-layer MLP that takes as input the concatenation of current state, local observations, actions, and task contextualization sampled from the joint task distribution, and outputs the next state, next local observations, and reward. The hidden dimension is 64 and the reconstruction loss is calculated by mean squared error. For (3) decentralized task approximation, the local trajectory encoders $f_{\theta_i^\prime} (i=1,\cdots,n)$ have the same structure as the global trajectory encoder $g_\theta$.

In benchmark Marines and SZ, a new difficulty arises since the number of agents, the dimension of state, observation, and action could vary from task to task, making the networks used in LBF and PP fail to work. Inspired by the popularly used population-invariant network (PIN) technique in MARL~\cite{hu2021updet,iqbal2021randomized,qin2022multi}, we design a different feature extractor, head and a monotonic mixing network~\cite{qmix} that learns the global Q-value as a combination of local Q-values. For the feature extractor (see Fig.~\ref{exp pin}), we decompose the observation $o_i$ into different parts, including agent $i$'s own information $o^{\rm own}_i$, ally information $o^{\rm al}_i$, and enemy information $o^{\rm en}_i$. Then we feed them into attention networks to derive a \textit{fixed-dimension} embedding $e$:
\begin{equation}
    \begin{aligned}
q & = \mathtt{MLP}_q( o^{\rm own}_i ),  \\
\quad\mathbf{K}_{\rm al} & = \mathtt{MLP}_{K_{\rm al}}([o^{{\rm al}_1}_i,\dots,o^{{\rm al}_j}_i,\dots]), \\ \mathbf{V}_{\rm al} & = \mathtt{MLP}_{V_{\rm al}}([o^{{\rm al}_1}_i,\dots,o^{{\rm al}_j}_i,\dots]),\\ 
e_{\rm al} & = \mathtt{softmax}(q\mathbf{K_{\rm al}}^ \mathrm{T}/\sqrt{d_k})\mathbf{V_{\rm al}}, \\
\quad\mathbf{K}_{\rm en} & = \mathtt{MLP}_{K_{\rm en}}([o^{{\rm en}_1}_i,\dots,o^{{\rm en}_j}_i,\dots]), \\ \mathbf{V}_{\rm en} & = \mathtt{MLP}_{V_{\rm en}}([o^{{\rm en}_1}_i,\dots,o^{{\rm en}_j}_i,\dots]),\\ 
e_{\rm en} & = \mathtt{softmax}(q\mathbf{K_{\rm en}}^ \mathrm{T}/\sqrt{d_k})\mathbf{V_{\rm en}}, \\
 e & = [ \mathtt{MLP}(o^{\rm own}_i), e_{\rm al} , e_{\rm en}],
\end{aligned}
\end{equation}
where $[\cdot, \cdot]$ is the vector concatenation operation, $d_k$ is the dimension of the query vector, and bold symbols are matrices. Embedding $e$ is then fed into a MLP and a GRU to derive the output of the feature extractor $\phi_i$. Finally, the output is fed into the policy head, a 3-layer MLP, to derive the Q-value. Furthermore, the dimension of states could also vary in Marines and SZ. Like the way we deal with observations, state $s$ is decomposed into ally information $s^{\rm al}_i$, and enemy information $s^{\rm en}_j$. Then their embeddings are fed into an attention network to derive a \textit{fixed-dimension} embedding $e_s$. Finally, we feed $e_s$ into the original mixing network whose structure is used in benchmarks LBF and PP.

Besides the networks mentioned above, the global trajectory encoder $g_\theta$, forward model $h$, local trajectory encoders $f_{\theta_i^\prime}$ are also involved with this issue. For $g_\theta$ and $f_{\theta_i^\prime}$, we first apply the same technique to derive the fixed-dimension embeddings of states and observations, then feed them into the transformer encoders. For the forward model $h$, we treat each agent's action as a part of its own observation and feed their concatenation into the attention network to derive an embedding, which will be feed into $h$ with the embedding of state and task contextualization. Then, $h$ outputs a fixed-dimension embedding. We decode it into the next state, local observations, and reward with task-specific MLP decoders to calculate the reconstruction loss $\mathcal{L}_{\text{model}}$.

\begin{algorithm}[!ht]
    \caption{MACPro: Training}
    \label{alg1}
    \textbf{Input}: Task sequence $\mathcal{Y} = \{\text{task}~1, \cdots, \text{task}~M\}$\\
    \textbf{Initialize}: trajectory encoder $g_\theta$, forward model $h$, individual trajectory encoders $f_{\theta_{i:n}^\prime}$, agents' feature extractors $\phi_{1:n}$
    \begin{algorithmic}[1] 
        \FOR{$m = 1, \cdots, M$}
            \STATE Set up task $m$
            \IF{$m = 1$}
                \STATE $\psi_{1:n}^{1} \leftarrow $ Initialized new head
            \ELSE
                \STATE \texttt{\small // Dynamic Network Expansion}
                \STATE Calculate $l, l^\prime$ according to Equation 5
                \STATE Find $k_* = \arg\min_{1 \le k \le K} l^\prime_{k}$
                \IF{$l^\prime_{k_*} \le \lambda_{\text{new}} l_{k_*}$}
                    \STATE Merge task $m$ to the task(s) that $\psi^{k_*}_{1:n}$ tasks charge, $\psi_{1:n}^{m} \leftarrow $ $\psi_{1:n}^{k_*}$
                \ELSE
                    \STATE $\psi_{1:n}^{m} \leftarrow $ Initialized new head
                \ENDIF
            \ENDIF
            \STATE (Optional) Reset the $\epsilon$-greedy schedule
            \FOR{$t = 1, \cdots, T_{\text{task}~m}$}
                \STATE Collect trajectories with $\{\phi_{1:n}, \psi_{1:n}^{m}\}$, store in buffers $\mathcal{D}, \mathcal{D}^\prime$
                \STATE Update $\{\phi_{1:n}, \psi_{1:n}^{m}\}$ according to $\mathcal{L}_{\text{RL}}$
                \IF{$t~mod~\kappa_1 = 0$}
                    \STATE \texttt{\small // Task Contextualization Learning}
                    \STATE Train $g_\theta, h$ according to $\mathcal{L}_{\text{context}}$
                    \STATE Train $f_{\theta_{1:n}^\prime}$ according to  $\mathcal{L}_{\text{approx}}$
                \ENDIF
                \IF{$t~mod~\kappa_2 = 0$}
                    \STATE Save head $\psi_{1:n}^m$ 
                \ENDIF
            \ENDFOR
            \STATE Evaluate task $1, \cdots, m$ according to Alg.~\ref{alg2}
            \STATE Empty the experience replay buffer $\mathcal{D} = \emptyset$
        \ENDFOR
    \end{algorithmic}
\end{algorithm}

\subsection{The Overall Flow of MACPro}
To illustrate the process of training, the overall training flow of MACPro is shown in Alg.~\ref{alg1}. Where Lines 3$\sim$14 express the process of dynamic agent network expansion, where we use the task contextualization to decide whether we should initialize a new head for the current task. To initialize a new head (Line 12), we propose two strategies. One strategy is to copy the parameters of the head learned from the last task. The other is to construct a entirely new head by resetting the parameters randomly. We use the first strategy in LBF, PP, Marines, and the second one in SZ. Both strategies work well in the experiments. Then, we started training on the new task. In Line 15 we can choose to reset the $\epsilon$-greedy schedule to enhance exploration (adopted in SZ) or not (adopted in LBF, PP, Marines), where the schedule is to decay $\epsilon$ from $1$ to $0.05$ in $50$K timesteps. Next, we iteratively update the parameters of each component in Lines 16$\sim$27, where we also save the current head. Finally, we test all seen tasks, empty the replay buffer for the current task, and switch to the next task.


Besides, the execution flow of MACPro is shown in Alg.~\ref{alg2}. In the execution phase, for each testing task, agents first toll-out $P$ episodes to probe the environment and derive the context (Line 3$\sim$7). With the gathered local information, each agent independently selects an optimal head to perform on this task (Line 8, 9).
\begin{algorithm}[!ht]
    \caption{MACPro: Execution}
    \label{alg2}
    \textbf{Input}: Task sequence $\{\text{task}~1, \cdots, \text{task}~M\}$, feature extractors $\phi_{1:n}$, heads $\{\psi_{1:n}^k\}_{k=1}^K$\\
    \textbf{Parameter}: Number of probing episodes $P$\\
    \begin{algorithmic}[1] 
        \FOR{$m = 1, \cdots, M$}
            \STATE Set up task $m$
            \FOR{$p = 1, \cdots, P$}
                \STATE Randomly choose an integer $k$ from $\{1, \cdots, K\}$
                \STATE Agents collect one trajectory $\pmb{\tau}_p$ with $\{\phi_{1:n}, \psi_{1:n}^{k}\}$
                \STATE Each agent $i$ calculates the mean value $\mu_{\theta_i^\prime} (\tau_p^i)$ of the trajectory representation $f_{\theta_i^\prime} (\tau_p^i)$
            \ENDFOR
            \STATE Each agent $i$ selects the optimal head $\psi_i^{{k^\star}^i}$, where ${k^\star}^i = \underset{1 \le k \le K}{\mathrm{argmin}}\, \underset{1 \le p \le P}{\mathrm{min}}\, ||\mu_{\theta_i^\prime}(\tau^i_p) - \frac{1}{bs} \sum\limits_{j=1}^{bs} \mu_k^j||_2.$
            \STATE Agents test with $\{\phi_i, \psi_i^{{k^\star}^i}\}_{i=1}^n$
        \ENDFOR
    \end{algorithmic}
\end{algorithm}

\begin{table}[!ht]
\footnotesize
\setlength{\tabcolsep}{0.9pt}
\renewcommand{\arraystretch}{1.3}
    \centering
    \caption{Hyperparameters in experiment.}
    \begin{tabular}{c|c}
        \toprule
        Hyperparameter  & Value \\
        \midrule
        Number of testing episodes & 32 \\
        $P$ (Number of probing episodes) & 20 \\
        Learning rate for updating networks & 0.0005 \\
        Number of heads in transformer encoders & 3 \\
        Number of layers in transformer encoders & 6 \\
        $bs$ (Batch size of the sampled trajectories) & 32 \\
        $\lambda_{\text{new}}$ (Threshold for merging similar tasks) & 1.5 \\
        Hidden Dim (Dimension of hidden layers) & 64 \\
        Attn Dim (Dimension of Key in attention) & 8 \\
        Entity Dim (Dimension of Value in attention) & 64 \\
        Z Dim (Dimension of the encoded Gaussians) & 32 \\
        $\kappa_1$ (Interval (steps) of updating both encoders) & 1000\\
        $\kappa_2$ (Interval (steps) of saving the learning head) & 10000\\
        $\alpha_{\text{cont}_l}$ (Coefficient of local contrastive loss $\mathcal{L}_{\text{cont}_l}$) & 0.1 \\
        $\alpha_{\text{cont}_g}$ (Coefficient of global contrastive loss $\mathcal{L}_{\text{cont}_g}$) & 0.1 \\
        $\alpha_{\text{reg}}$ (Coefficient of the $l_2$- regularization $\mathcal{L}_{\text{reg}}$ on $\phi$) & 500 \\
        Buffer Size of $\mathcal{D}$ (Maximum number of trajectories in $\mathcal{D}$) & 5000 \\
        Buffer Size of $\mathcal{D}^\prime$ (Maximum number of trajectories in $\mathcal{D}^\prime$) & 5000 \\
        \bottomrule
    \end{tabular}
    \label{hyperparameters}
\end{table}

\subsection{Hyperparameters Choices} 
Our implementation of MACPro is based on the PyMARL\footnote{\url{https://github.com/oxwhirl/pymarl}} \cite{pymarl} codebase with StarCraft 2.4.6.2.69232 and uses its default hyper-parameter settings. For example, the discounted factor used to calculate the temporal difference error is set to the default value 0.99.
The selection of the additional hyperparameters introduced in our approach, e.g., time interval of saving the heads, is listed in Tab.~\ref{hyperparameters}. We use this set of parameters in all experiments shown in this paper except for the ablations.

\section{The Complete Continual Learning Results} 
\label{moreresult}
\begin{figure*}
\setlength{\abovecaptionskip}{0cm}
  \centering
  \includegraphics[scale=0.50]{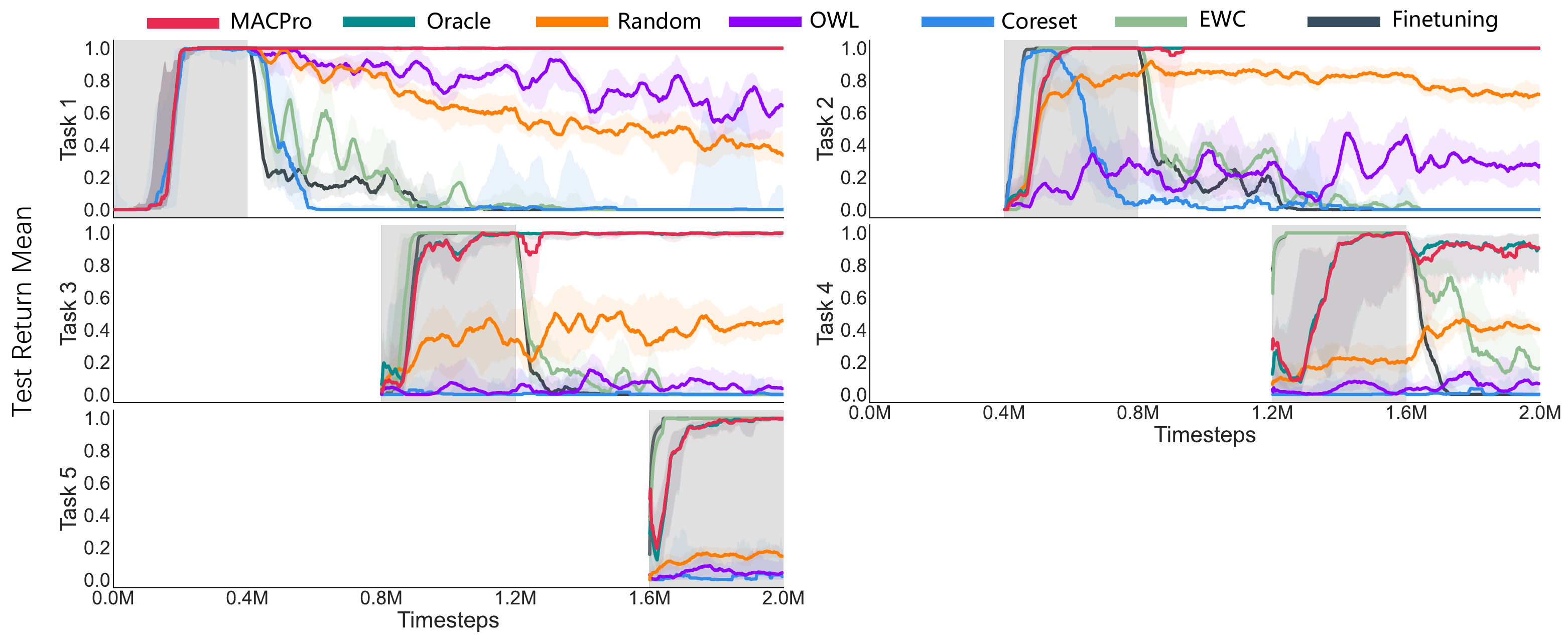}
  \caption{The complete continual learning results on LBF.}
  \label{exp lbf}
\end{figure*}

\begin{figure*}
\setlength{\abovecaptionskip}{0cm}
  \centering
  \includegraphics[scale=0.50]{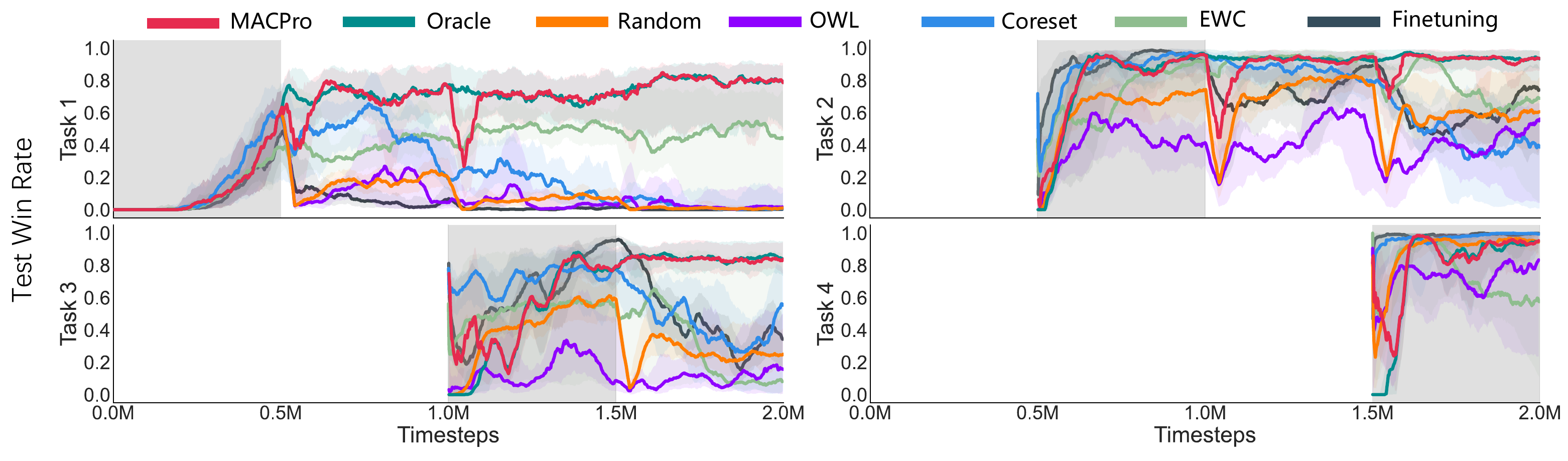}
  \caption{The complete continual learning results on SZ.}
  \label{exp sz}
\end{figure*}

In this part, we compare MACPro against the multiple mentioned baselines and ablations to investigate the continual learning ability, and display the performance on every single task seen so far in Fig.~\ref{exp lbf} $\sim$ \ref{exp sz}.

\begin{figure*}
\setlength{\abovecaptionskip}{0cm}
  \centering
  \includegraphics[scale=0.50]{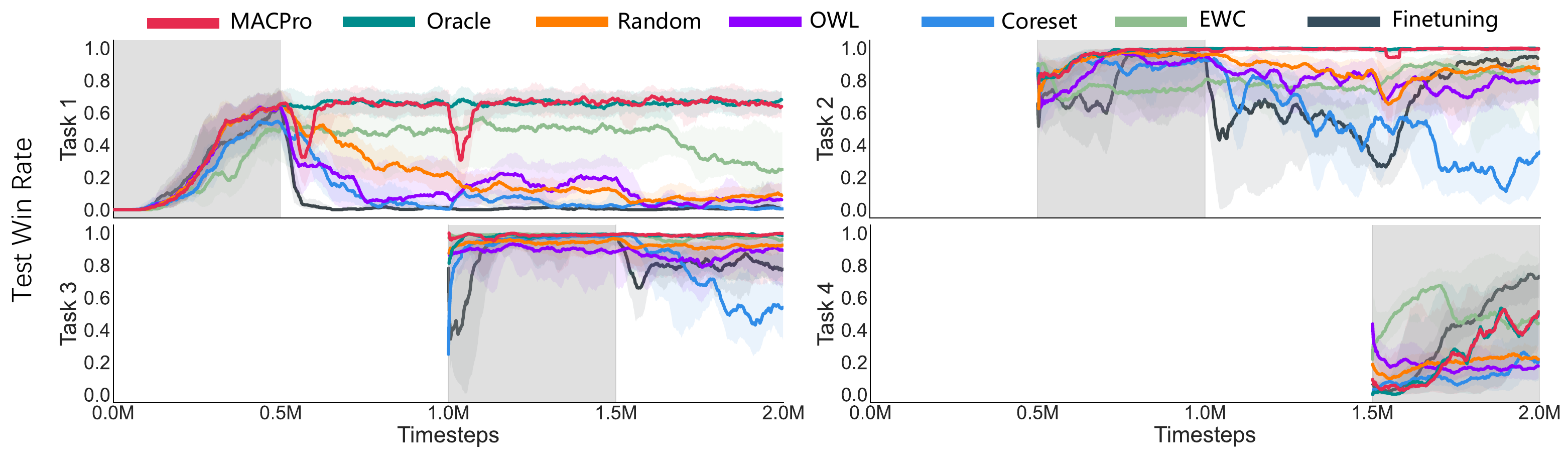}
  \caption{The complete continual learning results on Marines.}
  \label{exp marines}
\end{figure*}

\section{Product of a Finite Number of Gaussians}
\label{poeprof}
Suppose we have $N$ Gaussian experts with means $\mu_{i1},\mu_{i2},\cdots,\mu_{iN}$ and variances $\sigma_{i1}^2,\sigma_{i2}^2,\cdots,\sigma_{iN}^2$, respectively. Thus the product distribution is still Gaussian with mean $\mu_i$ and variance $\sigma_i^2$:
\begin{equation} \label{eqn.qwe}
\begin{aligned}
    \mu_i &= \Big(\frac{\mu_{i1}}{\sigma_{i1}^2}+\frac{\mu_{i2}}{\sigma_{i2}^2}+\cdots+\frac{\mu_{iN}}{\sigma_{iN}^2}\Big)\sigma_i^2, \\
    \frac{1}{\sigma_i^2} &= \frac{1}{\sigma_{i1}^2} + \frac{1}{\sigma_{i2}^2} + \cdots + \frac{1}{\sigma_{iN}^2}.
\end{aligned}
\end{equation}
It can be proved by induction.

\begin{proof}

We want to prove Eqn.~\ref{eqn.qwe} is true for all $N\ge 2$.

\begin{itemize}
    \item Base case: Suppose $N=2$ and $p_1(x)=\mathcal{N}(x|\mu_1,\sigma_1),p_2(x)=\mathcal{N}(x|\mu_2,\sigma_2)$, then
   \begin{equation} \label{eqn.p1}
\begin{aligned}
        & p_1(x)p_2(x) = \\
        & \frac{1}{\sqrt{2\pi}\sigma_1}\exp{\left(-\frac{\left(x-\mu_1\right)^2}{2\sigma_1^2}\right)}\cdot\frac{1}{\sqrt{2\pi}\sigma_2}\exp{\left(-\frac{\left(x-\mu_2\right)^2}{2\sigma_2^2}\right)} \\
        &= \frac{1}{2\pi\sigma_1\sigma_2}\exp{\left(-\left(\frac{\left(x-\mu_1\right)^2}{2\sigma_1^2}+\frac{(x-\mu_2)^2}{2\sigma_2^2}\right)\right)}  \\
        &= \frac{1}{2\pi\sigma_1\sigma_2}\exp{\left(-\frac{x^2-2\frac{\mu_1\sigma_2^2+\mu_2\sigma_1^2}{\sigma_1^2+\sigma_2^2}x+\frac{\mu_1^2\sigma_2^2+\mu_2^2\sigma_1^2}{\sigma_1^2+\sigma_2^2}}{2\frac{\sigma_1^2\sigma_2^2}{\sigma_1^2+\sigma_2^2}}\right)} \\
        &= \frac{1}{2\pi\sigma_1\sigma_2}\exp{\left(-\frac{\left(x-\frac{\mu_1\sigma_2^2+\mu_2\sigma_1^2}{\sigma_1^2+\sigma_2^2}\right)^2}{2\frac{\sigma_1^2\sigma_2^2}{\sigma_1^2+\sigma_2^2}}-\frac{\left(\mu_1-\mu_2\right)^2}{2\sigma_1^2\sigma_2^2}\right)}\\ 
        &= \frac{1}{\sqrt{2 \pi\left(\sigma_{1}^{2}+\sigma_{2}^{2}\right)}}\exp{\left(-\frac{\left(\mu_{1}-\mu_{2}\right)^{2}}{2\left(\sigma_{1}^{2}+\sigma_{2}^{2}\right)}\right)}\\ & \cdot \frac{1}{\sqrt{2 \pi} \frac{\sigma_{1} \sigma_{2}}{\sqrt{\sigma_{1}^{2}+\sigma_{2}^{2}}}} 
         \cdot \exp{\left(-\frac{\left(x-\frac{\mu_{1} \sigma_{2}^{2}+\mu_{2} \sigma_{1}^{2}}{\sigma_{1}^{2}+\sigma_{2}^{2}}\right)^{2}}{2 \frac{\sigma_{1}^{2} \sigma_{2}^{2}}{\sigma_{1}^{2}+\sigma_{2}^{2}}}\right)}\\
        &= A\cdot \frac{1}{\sqrt{2 \pi} \frac{\sigma_{1} \sigma_{2}}{\sqrt{\sigma_{1}^{2}+\sigma_{2}^{2}}}} \exp\left(-\frac{\left(x-\frac{\mu_{1} \sigma_{2}^{2}+\mu_{2} \sigma_{1}^{2}}{\sigma_{1}^{2}+\sigma_{2}^{2}}\right)^{2}}{2 \frac{\sigma_{1}^{2} \sigma_{2}^{2}}{\sigma_{1}^{2}+\sigma_{2}^{2}}}\right). 
\end{aligned}
 \end{equation}
    Eqn.~\ref{eqn.p1} can be seen as PDF of $\mathcal{N}(\mu,\sigma)$  times $A$ where $\mu = (\frac{\mu_1}{\sigma_1^2}+\frac{\mu_2}{\sigma_2^2})\sigma^2,
    \frac{1}{\sigma^2} = \frac{1}{\sigma_1^2} + \frac{1}{\sigma_2^2}.$
    
    \item Induction step: Suppose it is true when $N=n$, and the product distribution of $n$ Gaussian experts has mean $\tilde{\mu}=(\frac{\mu_1}{\sigma_1^2}+\cdots+\frac{\mu_n}{\sigma_n^2})\tilde{\sigma}^2$ and variance $\frac{1}{\tilde{\sigma}^2}=\frac{1}{\sigma_1^2} + \cdots+\frac{1}{\sigma_n^2}$, then for $n+1$ Gaussian experts:
       \begin{equation} 
\begin{aligned}
       & \frac{1}{\sigma^2} = \frac{1}{\tilde{\sigma}^2} + \frac{1}{\sigma_{n+1}^2} = \frac{1}{\sigma_1^2} + \cdots+\frac{1}{\sigma_n^2}+ \frac{1}{\sigma_{n+1}^2},\\
        &\mu = \Big(\frac{\tilde{\mu}}{\tilde{\sigma}^2}+\frac{\mu_{n+1}}{\sigma_{n+1}^2}\Big)\sigma^2 = \Big(\frac{\mu_1}{\sigma_1^2}+\cdots+\frac{\mu_n}{\sigma_n^2}+\frac{\mu_{n+1}}{\sigma_{n+1}^2}\Big)\sigma^2.
 \end{aligned}
 \end{equation}
    
    \item Eqn.~\ref{eqn.qwe} has been proved by the above derivation.
\end{itemize} 
\end{proof}

If we write $T_{ij}=(\sigma_{ij}^2)^{-1}$,  Eqn.~\ref{eqn.qwe} can be written as:
       \begin{equation} 
\begin{aligned}
    \mu_i &= \Big(\sum_{j=1}^N \mu_{ij}T_{ij}\Big)\Big(\sum_{j=1}^N T_{ij}\Big)^{-1}, \\
    \sigma_i^2 &= \Big(\sum_{j=1}^N T_{ij}\Big)^{-1},
 \end{aligned}
 \end{equation}
and is exactly what we're trying to prove.

\end{document}